\documentclass[a4paper,11pt]{article}
\pdfoutput=1

\usepackage{jcappub}
\usepackage[utf8]{inputenc}

\title{Inclusive Displaced Vertex Searches for Heavy Neutral Leptons at the LHC}
\author[a]{Asmaa Abada,}
\author[b]{Nicol\'as Bernal,}
\author[b]{Marta Losada}
\author[a]{and\\Xabier Marcano}
\affiliation[a]{Laboratoire de Physique Th\'eorique, CNRS\\
Univ. Paris-Sud, Universit\'e Paris-Saclay, 91405 Orsay, France}
\affiliation[b]{Centro de Investigaciones, Universidad Antonio Nari\~no\\
Carrera 3 Este \# 47A-15, Bogot\'a, Colombia}
\emailAdd{abada@th.u-psud.fr}
\emailAdd{nicolas.bernal@uan.edu.co}
\emailAdd{malosada@uan.edu.co}
\emailAdd{xabier.marcano@th.u-psud.fr}

\abstract{
	The inclusion of heavy neutral leptons  to the Standard Model particle content could provide solutions to many open questions in particle physics and cosmology. The modification of the charged and neutral currents from active-sterile mixing of neutral leptons can provide novel signatures in Standard Model processes. 
We revisit the displaced vertex signature that could occur in collisions at the LHC via the decay of heavy neutral leptons   with masses of a few GeV emphasizing the implications of flavor,  kinematics, inclusive production and number of these extra neutral fermions.
We study in particular the implication on the parameter space sensitivity  when all mixings to active flavors are taken into account. We also discuss alternative cases  where the new particles  are produced in  a  boosted  regime.
}

\begin{document}
\begin{flushright}
	LPT-ORSAY-18-79\\
	PI/UAN-2018-631FT
\end{flushright}
\maketitle


\section{Introduction}

A natural and simple extension of the Standard Model  (SM) accommodating theoretical and observational challenges calls upon the existence of neutral leptons which are sterile from the SM gauge point of view (see Ref.~\cite{Abazajian:2012ys} for a comprehensive review). Their unique interaction with the SM particles is via their mixing with the active neutrinos (Yukawa interaction). There is {\it a priori} no limit on the mass scale associated to these extra fermions. 

In this work we focus on collider experimental tests of their existence, in particular in the mass range up to a few tens of GeV, without invoking a link to a specific neutrino mass generation mechanism of these extra neutral fermions (dubbed `{\it heavy neutral leptons}',  HNL).
Many experimental searches have focused on the high mass regime (including masses above the $W$ boson mass) where the HNL are  produced directly or in some  prompt decay channels~\cite{Abreu:1996pa,Aad:2011vj,Chatrchyan:2012fla,Khachatryan:2015gha,Aad:2015xaa,
Khachatryan:2016olu,Sirunyan:2018mtv}, with numerous dedicated  analysis, for instance in Refs.~\cite{delAguila:2007qnc,delAguila:2008cj,delAguila:2008hw,Atre:2009rg,Dev:2013wba,Das:2014jxa,Alva:2014gxa,Deppisch:2015qwa,Banerjee:2015gca,Arganda:2015ija,Das:2015toa,Degrande:2016aje,Mitra:2016kov,Ruiz:2017yyf,Cai:2017mow,Pascoli:2018rsg}.
Having relatively light heavy neutral fermions that do not decouple since they could have sufficiently large  mixings  with active SM neutrinos may lead to important consequences, a major one being the modification of the charged and neutral currents with a leptonic mixing matrix (encoding the PMNS mixing matrix~\cite{Pontecorvo:1957cp,Maki:1962mu} and the active-HNL mixings). Moreover, and if sufficiently light, the HNL can be produced as final states. Both of these two features  impact several observables, leading  at the same time to abundant constraints on  parameter space, i.e. mixing angles and masses of the HNL (see Refs.~\cite{Atre:2009rg,Abada:2017jjx} and references therein).

An exciting possibility occurs in a region of parameter space for which the HNL are long-lived particles that can decay with a sizable displacement in the LHC detectors, 
and such displaced vertices (DV) would be a distinctive signature of their existence~\cite{Gronau:1984ct}. 
Recently, several studies have pushed for these kinds of dedicated LHC searches for HNL with associated charged leptons~\cite{Nemevsek:2011hz,Helo:2013esa,Izaguirre:2015pga,Dube:2017jgo,Cottin:2018kmq,Cottin:2018nms,Dib:2018iyr,Nemevsek:2018bbt}, from Higgs decays~\cite{Maiezza:2015lza,Gago:2015vma,Accomando:2016rpc,Nemevsek:2016enw,Caputo:2017pit,Deppisch:2018eth,Liu:2018wte}, for LHCb~\cite{Antusch:2017hhu} or at future detectors proposed for searching for long-lived particles~\cite{Caputo:2016ojx,Kling:2018wct,Helo:2018qej,Jana:2018rdf,Curtin:2018mvb}.
There is also a potential to search for such signatures at DUNE~\cite{Adams:2013qkq}, Icecube~\cite{Coloma:2017ppo}, future lepton colliders~\cite{Blondel:2014bra,Antusch:2016vyf} and SHiP~\cite{Alekhin:2015byh,Anelli:2015pba}, the latter expected to highly improve sensitivity to HNL below the charm quark mass, with the HNL being abundantly produced by meson decays~\cite{Bonivento:2013jag}.

At the LHC, the cross sections of the single production of  $W^\pm$, $Z$ and $H$ bosons have been measured~\cite{CMS:2015ois,Aad:2016naf,ATLAS:2017ovn,CMS:2018lkl} and given the expected  integrated luminosity,  it would be possible to have a large number of events  in which these bosons decay  into a HNL. In addition, the LHC has successfully measured a large number of  SM processes such as  diboson,  $t \overline{t}$, $W/Z+$ jets, etc. 
As mentioned above, the modification of the charged and neutral currents due to the active-sterile mixings opens up additional decay channels for these SM processes occurring at the LHC, in which a $W^\pm$, $Z$ and the $H$ boson will decay.

In this work we address the DV signature in  two  types of scenarios: $i)$ large number of events like the Drell-Yan processes, $ii)$ other SM processes with a smaller cross section mostly relevant for the high luminosity phase of the LHC. In both cases we look at the DV signature possibility focusing on the production mechanisms of  the HNL when they are long-lived and decay in the detector. 

In the first case, we consider the {\it inclusive production}   from Drell-Yan, $W^\pm$,  $Z$ and $H$ decay processes   for which 
we do not apply selection cuts on the primary vertex producing the HNL, but instead focus on the signature of the DV from the HNL decay. This allows  a larger number of total events. The  experiments at the LHC will be sensitive for a given region of the parameter space (mixings and masses of the HNL) that we identify in this work. 

We also exemplify in this work the flavor dependence for a particular production and decay channel given that the decay width is proportional to the sum of the square of the mixings, implying that all these channels are complementary and necessary to probe the whole parameter space (masses of the HNL and their mixings to the SM active neutrinos); for this we consider different cases of the active-sterile mixing pattern.  This is in contrast with what has been done for instance in Refs.~\cite{Helo:2013esa,Izaguirre:2015pga,Dube:2017jgo,Cottin:2018kmq,Cottin:2018nms}, where only one flavor was considered.

In the second case, we consider other SM processes in which the gauge bosons are produced in combination with other particles. In this case, the gauge bosons might be boosted and the kinematic of their decay products might be  significantly different from that of the Drell-Yan processes. In particular, for those events in which the HNL is strongly boosted, distinctive DV signatures can take place. We illustrate in this work the potential of this DV signature for a specific process.

Throughout the study, we do not consider a specific neutrino mass generation mechanism but rather consider a bottom-up approach extending the SM with ${\mathcal{N}}$ sterile fermions. Most of the analysis is conducted for ${\mathcal{N}}=1$ and extended to ${\mathcal{N}}=2$ to  show (new) distinctive features of DV signatures compared to the case  with $\mathcal{N}=1$.

This work is organized as follows: in Section~\ref{sec:framework} we present the framework we adopt that is the simple  extension of the SM via sterile fermions, $3+{\mathcal{N}}$ and provide details of the numerical simulations we perform. 
Section~\ref{sec:HNLdecay} is devoted to the determination of the HNL decay width discussing the different HNL decay channels and highlighting the role of flavor. 
We present and discuss the results of DV arising from inclusive HNL production at the LHC in Section~\ref{sec:inclusive}, considering both flavored and flavor blind cases for the production of HNL. Furthermore,  we discuss in detail the LHC sensitivities for the dimuon channel as a specific case. We also extend the study to the  $3+2$ model (${\mathcal{N}}=2$). Finally, we discuss in Section~\ref{sec:boosted} the results we obtain in a specific SM process in which the HNL is strongly boosted ($W/Z+$ jets).
Our results are summarized in Section~\ref{sec:conclusion}. 
 The parametrization we use for the $3+ {\mathcal{N}}$ model (${\mathcal{N}}=1,\,2$) is given in Appendix~\ref{app:sec:parametrization}.

\section{Theoretical Framework}\label{sec:framework}

In this work we are interested in studying  collider phenomenology of HNLs without assuming any specific underlying model or mechanism of light neutrino mass generation.
In order to do this, we follow a bottom-up approach where the SM is extended by {\it ad-hoc} masses for the 3 active neutrinos, as required by  oscillation phenomena, and by $\mathcal{N}$ additional sterile fermions (the HNL). 
We refer to these kinds of scenarios as $3+\mathcal{N}$ models, and they are useful to understand the general phenomenology of a broad class of models where the SM is enlarged only by sterile fermions, such as the type-I seesaw model and its variants.

In most of our forthcoming analysis, we will focus on the simplest case of having only one HNL ($\mathcal{N}=1$), although we will discuss the generalization to the case with more HNLs in Section~\ref{sec2HNL}.
We will also assume that neutrinos are Majorana fermions.\footnote{In this study we did not study lepton flavor violating processes.}
However, it is worth stressing that the Majorana nature is linked to lepton number violation and in this minimal framework, one must assume a link between the smallness of the active neutrino masses and an approximate lepton number conservation. Although this study concerns collider searches, it was recently shown that lepton number-violating currents involving only a type-I Seesaw field content will be very suppressed in collider experiments~\cite{Moffat:2017feq}, making the production of Majorana neutrinos very small. One must then assume some variant of the type-I seesaw with an approximate lepton number symmetry, such as the inverse~\cite{Wyler:1982dd,Mohapatra:1986bd,GonzalezGarcia:1988rw,Abada:2014vea}, and/or linear seesaw mechanisms~\cite{Barr:2003nn,Malinsky:2005bi} in which one could have a seesaw realization for the active neutrino masses at low scale and comparatively large Yukawa couplings. Another possibility would be that the gauge group of the SM is enlarged so that 
 Majorana neutrino's production is not through the electroweak current, like in the case of the left-right symmetric model (see e.g. Ref.~\cite{Ruiz:2017nip}).
In the following, we consider only the physical masses of the HNL and their mixings to the active neutrino sector without specifying their origin. 

The $3+1$ model consists of four neutrino masses
and a unitary $4\times4$ leptonic mixing matrix,
\begin{equation}
U_\nu^{3+1} = \left(\begin{array}{cc}
\tilde U_{\rm PMNS} & V_{\ell N}\\
V_{N\ell} & U_{NN}
\end{array}
\right)\,,
\end{equation}
where the $3\times3$ matrix $\tilde U_{\rm PMNS}$ is the usual PMNS matrix up to non-unitarity corrections due to the presence of light-sterile $V_{\ell N}$ mixings, with $\ell=e$, $\mu$, $\tau$.
These mixings, together with the HNL mass $m_N$, define the interaction strength of the HNL via charged currents, as well as the neutral currents to both $Z$ and $H$ bosons.
Therefore, they will be the relevant parameters for our study. 
In the case of an extension of the SM with $\mathcal{N}$ HNL, the relevant terms in the Lagrangian (in the Feynman-'t Hooft gauge)  are  given below:
\begin{eqnarray}
\mathcal{L}&=&
-\frac{g}{\sqrt{2}}U_{\alpha
 i}W_\mu^{-}\overline{\ell_\alpha}\gamma^{\mu}P_L\nu_i
-\frac{g}{\sqrt{2}}U_{\alpha
i}H^{-}\overline{\ell_\alpha}\left(\frac{m_\alpha}{m_W}P_L-\frac{m_i}{m_W}P_R\right)\nu_i 
+\mathrm{H.c.}\nonumber\\
&&
-\frac{g}{2\cos\theta_W}U_{\alpha i}^*U_{\alpha
j}Z_{\mu}\overline{\nu_i}P_L\nu_j
-\frac{ig}{2}U_{\alpha i}^*U_{\alpha j}A^0\overline{\nu_i}\left(\frac{m_j}{m_W}P_R\right)\nu_j
+\mathrm{H.c.}\nonumber\\
&&
-\frac{g}{2}U_{\alpha i}^*U_{\alpha
j}h\overline{\nu_i}\left(\frac{m_j}{m_W}P_R\right)\nu_j
+\mathrm{H.c.}\ , 
\label{eq:lag}
\end{eqnarray}
where $g$ is the $SU(2)_L$ gauge coupling, $U_{\alpha i}\equiv U_{\nu_{\alpha i}}^{3+\mathcal{N}}  $ are the 
 lepton mixing matrix components, 
$m_i$  are the mass eigenvalues of the neutrinos and  $m_\alpha$ are the charged lepton masses. 
The indices $\alpha$ and $i$ run as  $\alpha=e$, $\mu$, $\tau$ and $i=1$\dots\,$3+\mathcal{N}$.
 
The existence of a HNL has been tested in different observables, depending on its mass, and at present there are strong bounds on its mixings (see Ref.~\cite{Abada:2017jjx} for a recent update). 
In the range of masses we will be interested on, i.e. the few GeV regime, the strongest upper bounds are those from DV searches by DELPHI~\cite{Abreu:1996pa}, which constrains the sum of the squared mixings to be below $2\times10^{-5}$.
We will consider these bounds in our numerical analysis and discuss how the LHC could improve them searching also for DV signatures.

For our forthcoming numerical simulations, we have implemented  the 3+1 and 3+2 models in {\tt FeynRules}~\cite{Christensen:2008py,Alloul:2013bka} to generate the UFO model file~\cite{Degrande:2011ua}, taking also into account the effective Higgs coupling to gluons for Higgs production. 
Then, we use  {\tt MadGraph5\_aMC@NLO}~\cite{Alwall:2014hca}  to generate  the HNL production events from proton-proton collisions and  {\tt Pythia 8.2}~\cite{Sjostrand:2014zea} for its subsequent decay and treatment of the DV. 
Finally, Les Houches Event files are obtained using {\tt MadAnalysis5}~\cite{Conte:2012fm}.
Jets are reconstructed using an anti-$k_T$ algorithm with a radius of 0.4 and minimum $p_T$ of 5~GeV.
The HNL lifetimes are computed analytically and included them in {\tt MadGraph5}, using the {\tt time\_of\_flight} option to generate the events with DV.

\section{Decays of the Heavy Neutral Lepton}
\label{sec:HNLdecay}

When exploring signatures from DV, the total width of the decaying particle is a crucial parameter. 
Generally speaking,  its decay length must be of the same order of the size of the detector, or more specifically of the tracker system, which in the case of detectors such as ATLAS or CMS cover  transverse displacements between roughly 1~mm and 1~m from the interaction point.
In this section we explore the decays of the HNL and discuss the parameter space  that could be explored by DV searches at the LHC.
 In the following, we will consider the Majorana HNL case, whose total width is twice the one in the Dirac case, since it can decay to both $CP$ conjugated final states. 
 
The decays of the HNL can be divided in different regimes, depending on its mass~\cite{Atre:2009rg,Abada:2017jjx,Bondarenko:2018ptm}. 
When the mass is below the GeV scale, it mainly decays   via off-shell $W$ or $Z$ bosons, leading to  three body leptonic or two body semileptonic final states, i.e. $N\to \ell\ell'\nu$, $N\to\nu\nu\nu$ or $N\to \ell M$, $M$ being a light meson. 
Above a GeV but still below the $W$ mass, the semileptonic decays are better estimated by the three body decay to quarks, $N\to \ell q\bar q'$, which also accounts  for the possible hadron multiproduction. 
Finally, if the HNL is very massive, $m_N>m_{W}$, it tends to decay very fast via on-shell $W$, $Z$ and $H$ bosons, and, thus, one needs to search for prompt decay signatures.

\begin{figure}[t!]
\begin{center}
\includegraphics[width=0.49\textwidth]{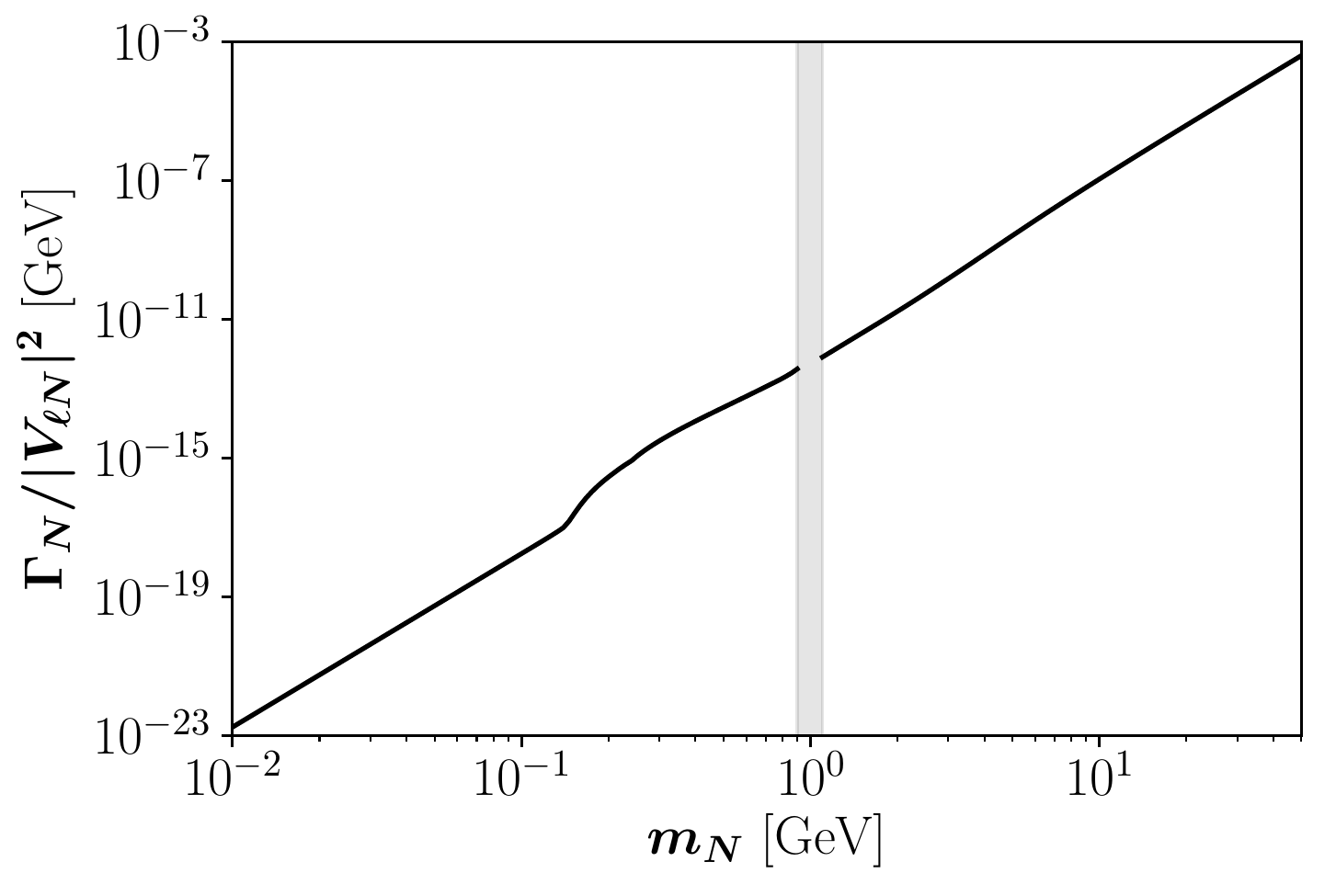}
\includegraphics[width=0.49\textwidth]{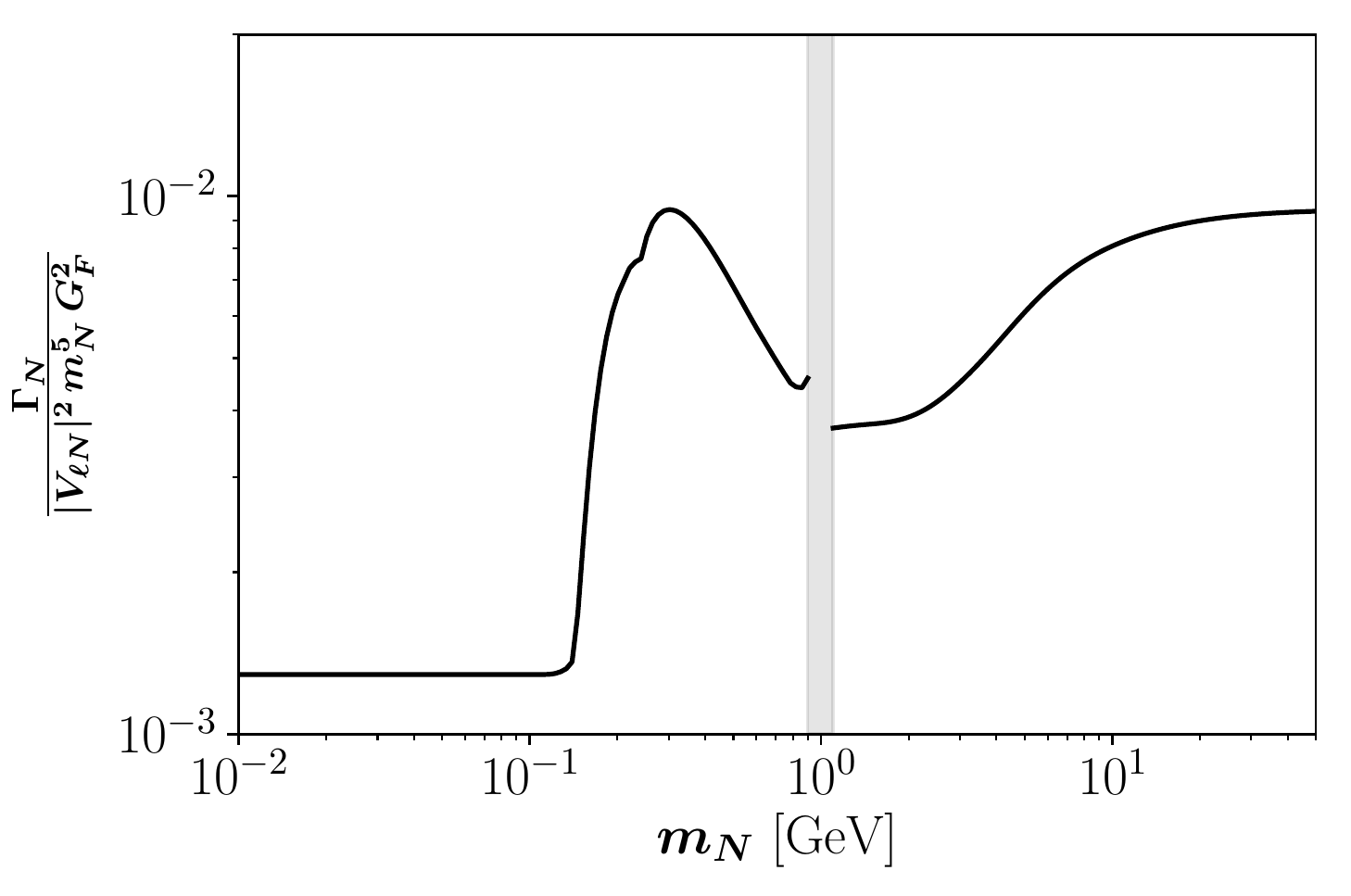}
\includegraphics[width=0.49\textwidth]{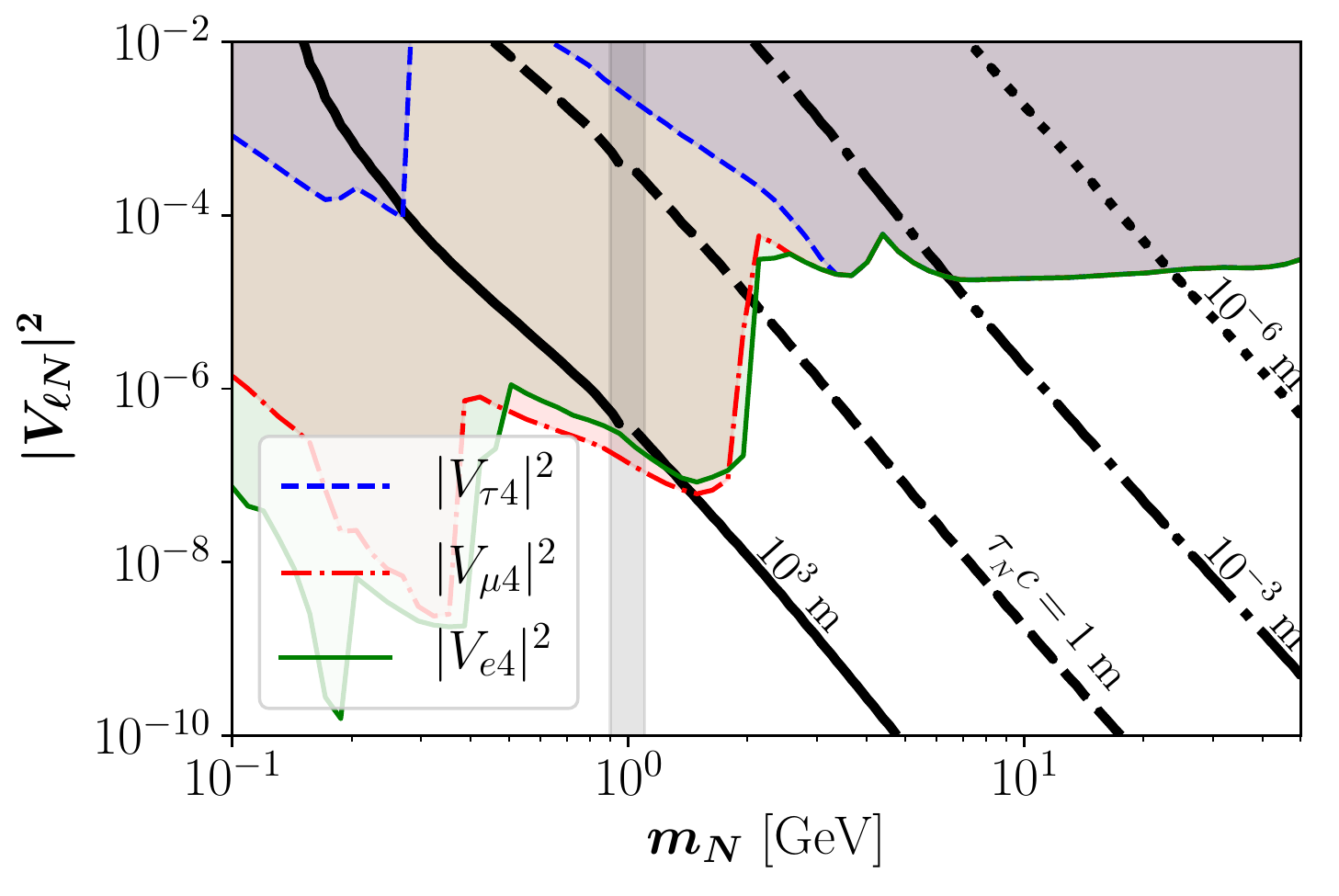}
\caption{
Upper panels: (left) HNL total decay width $\Gamma_N$  as a function of its mass $m_N$ and normalized to the mixings for $|V_{eN}|=|V_{\mu N}|=|V_{\tau N}|\equiv |V_{\ell N}|$; (right) the same normalized to $m_N^5\,G_F^2$.
Lower panel: contour lines for the HNL decay length $\tau_{_N} c$. Shaded areas are excluded by different experiments.
}\label{Decay_mN}
\end{center}
\end{figure}

We show in Fig.~\ref{Decay_mN} the total neutrino width $\Gamma_N$ (upper panels) and the decay length $\tau_{_N} c$ (lower panel) as a function of the HNL mass $m_N$ and mixing $|V_{\ell N}|$ in a ``democratic" scenario of $|V_{eN}|=|V_{\mu N}|=|V_{\tau N}|$. 
The vertical gray lines are only to illustrate the transition between the semileptonic decay to mesons, $N\to \ell M$, and to quarks, $N\to \ell q\bar q'$. 
In the upper panels we see how the increase with $m_N$ and $|V_{\ell N}|$ is translated into the diagonal contour lines in the lower one.
The black lines in the latter show contours for $\tau_{_N} c=10^3$, 1, $10^{-3}$ and $10^{-6}$~m, while shaded areas are experimentally excluded~\cite{Abada:2017jjx}.
Interestingly, the region with decay lengths relevant for DV searches at the LHC, say $\tau_{_N} c\in$~[1~mm, 1~m], lies in the few GeV region, where present experimental constraints on the mixing angles are weaker.
In a general case of three different mixing angles, we have checked that the behavior of the total width  can be approximated as
\begin{equation}\label{eq:Nwidth}
\Gamma_N\propto G_F^2\, m_N^5 \sum_{\ell=e,\,\mu,\,\tau} \big| V_{\ell N} \big|^2\,,
\end{equation}
which works very well within this area of interest for the DV at LHC, especially above the tau mass threshold.
For the numerical analysis, the total decay widths and the branching ratios of the HNL were computed analytically and used as inputs for {\tt MadGraph5\_aMC@NLO}.

At this point, a remark about the role of flavor  is in order. 
In many collider searches for HNL, one often considers simplified hypothesis where the HNL mixes only to one flavor at a time. 
This simplification is well justified in prompt decayed HNL searches with dilepton or trilepton final states, since the number of events with charged leptons of a flavor $\ell$ depends only on the corresponding mixing $V_{\ell N}$. 
Nevertheless, this does not apply to the DV searches.
In this case, the total decay width plays a crucial role defining where the HNL will decay and, since it depends on all the mixings $V_{\ell N}, \ell=e,\mu,\tau$, one cannot conclude independently of each of the mixings in DV searches.

\begin{figure}[t!]
	\begin{center}
	\includegraphics[width=.65\textwidth]{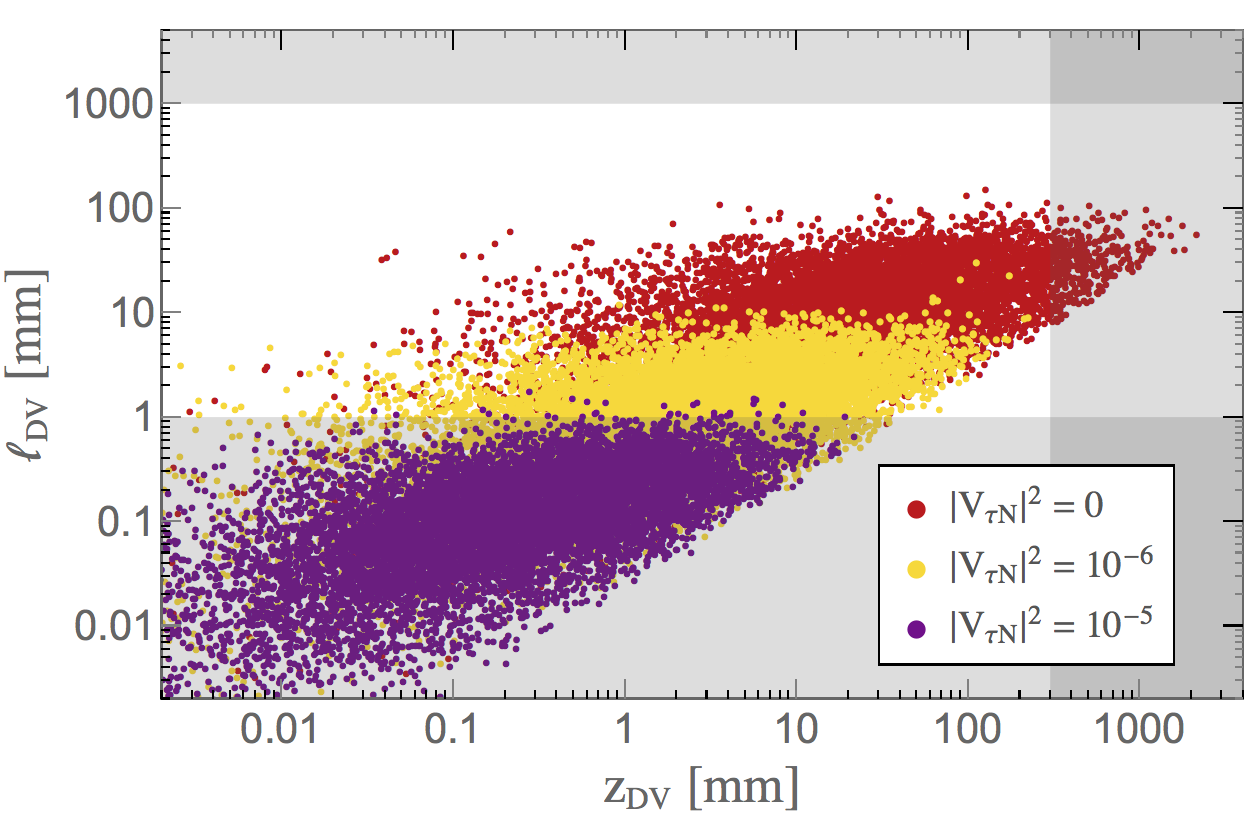}
		\caption{Distributions of the DV coming from a HNL decay in the $(z_\text{DV}, \ell_\text{DV})$ plane. The production channel is $pp\to e^\pm N$, with $p_T^e>25$GeV and $ |\eta^e|<2.5$. We fix $m_N=15$~GeV, $|V_{eN}|^2=10^{-7}$ and $V_{\mu N}=0$ in all the cases, while $|V_{\tau N}|^2= 0$ (red), $10^{-6}$ (yellow) and $10^{-5}$ (purple). As $|V_{\tau N}|^2$ increases, the HNL becomes more prompt and, therefore, insensitive to the DV searches. The white region is the potential  DV area.
	}\label{DVdistributionsVtauN}
	\end{center}
\end{figure}

In order to illustrate this effect, we display in Fig.~\ref{DVdistributionsVtauN} the distributions for the HNL decay position in the $(z_{\rm DV},\,\ell_{\rm DV})$ plane, where $z_{\rm DV}$ is the displacement along the beam axis and $\ell_{\rm DV}$ in the transverse plane.
We have generated $pp\to e^\pm N$ events with fixed values of $m_N=15$~GeV, $|V_{eN}|^2=10^{-7}$ and $V_{\mu N}=0$, and take the $|V_{\tau N}|^2$ mixing equal to zero (red), $10^{-6}$ (yellow) and $10^{-5}$ (purple). 
Since the mass and electron mixing are fixed, the same number of HNL are produced in all  cases.
Nevertheless, the sensitivity of the DV searches to the HNL  depends on the amount of events within the DV fiducial volume, $\ell_{\rm DV} \in [1~{\rm mm},\, 1~{\rm m}]$ and $z_\text{DV}<300$~mm. 
In this example, these kinds of searches are very sensitive to the single mixing case, as  most of the red points are within this area.
However, increasing values of $V_{\tau N}$ enhances the total width without affecting the production cross section, shifting the distribution towards lower displacements and thus reducing  the efficiency of the DV searches. Alternatively, some of the cases escaping the detector in the single mixing scenario could lead to DV signatures when all mixings are taken into account. 
Consequently, these kinds of searches cannot  explore independently the mixing of each flavor and they should take into account the flavor combination entering in the total width in Eq.~\eqref{eq:Nwidth}.
We will discuss the impact of this effect on the final sensitivity estimates in Section~\ref{sec:inclusive}.

\begin{figure}[t!]
        \begin{center}
        \includegraphics[width=.32\textwidth]{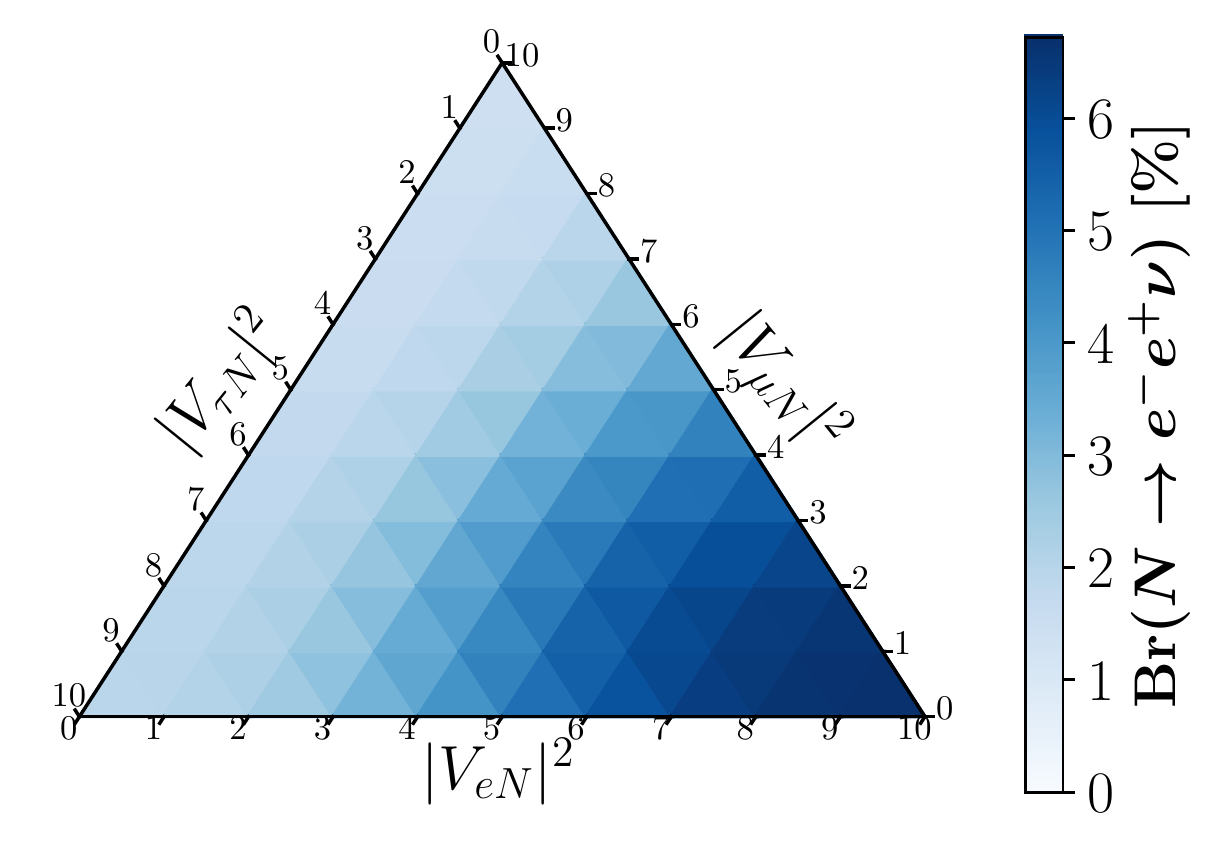}
        \includegraphics[width=.32\textwidth]{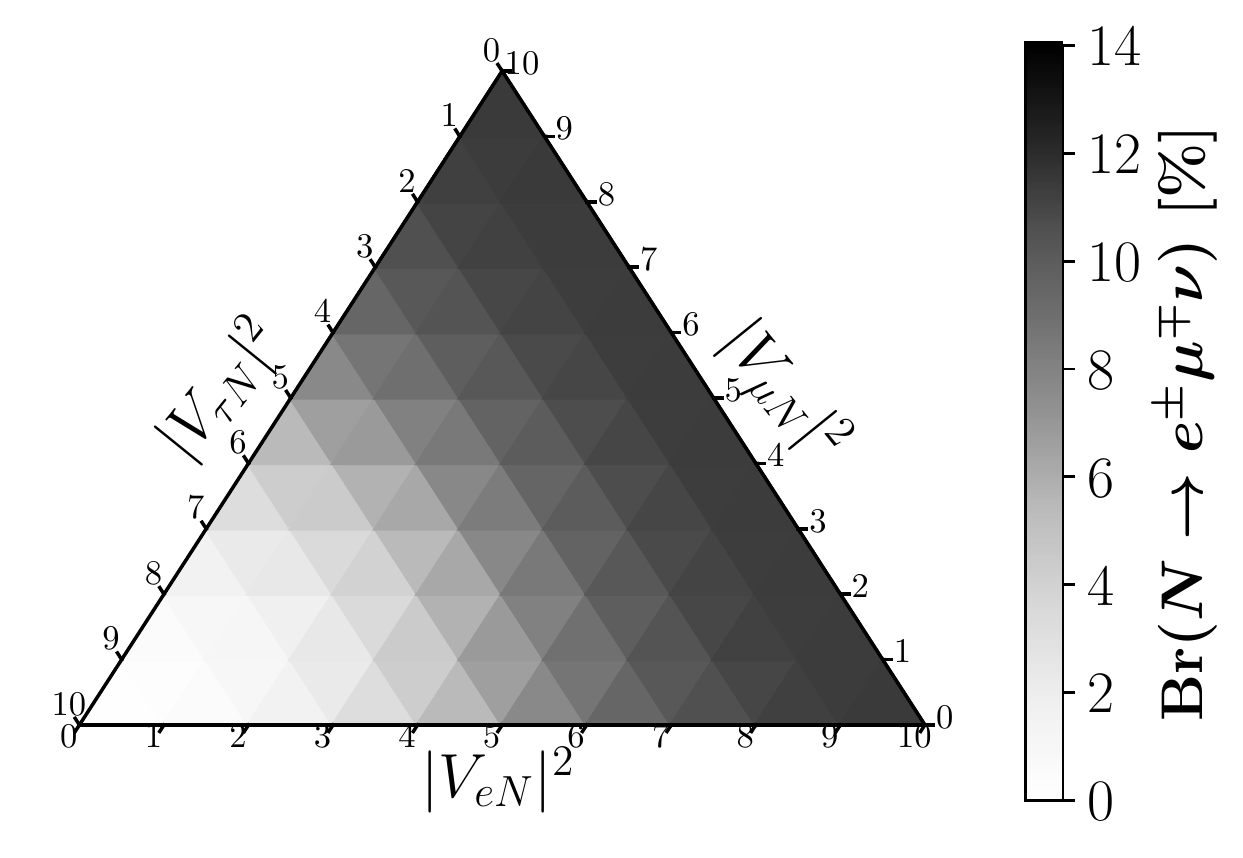}
        \includegraphics[width=.32\textwidth]{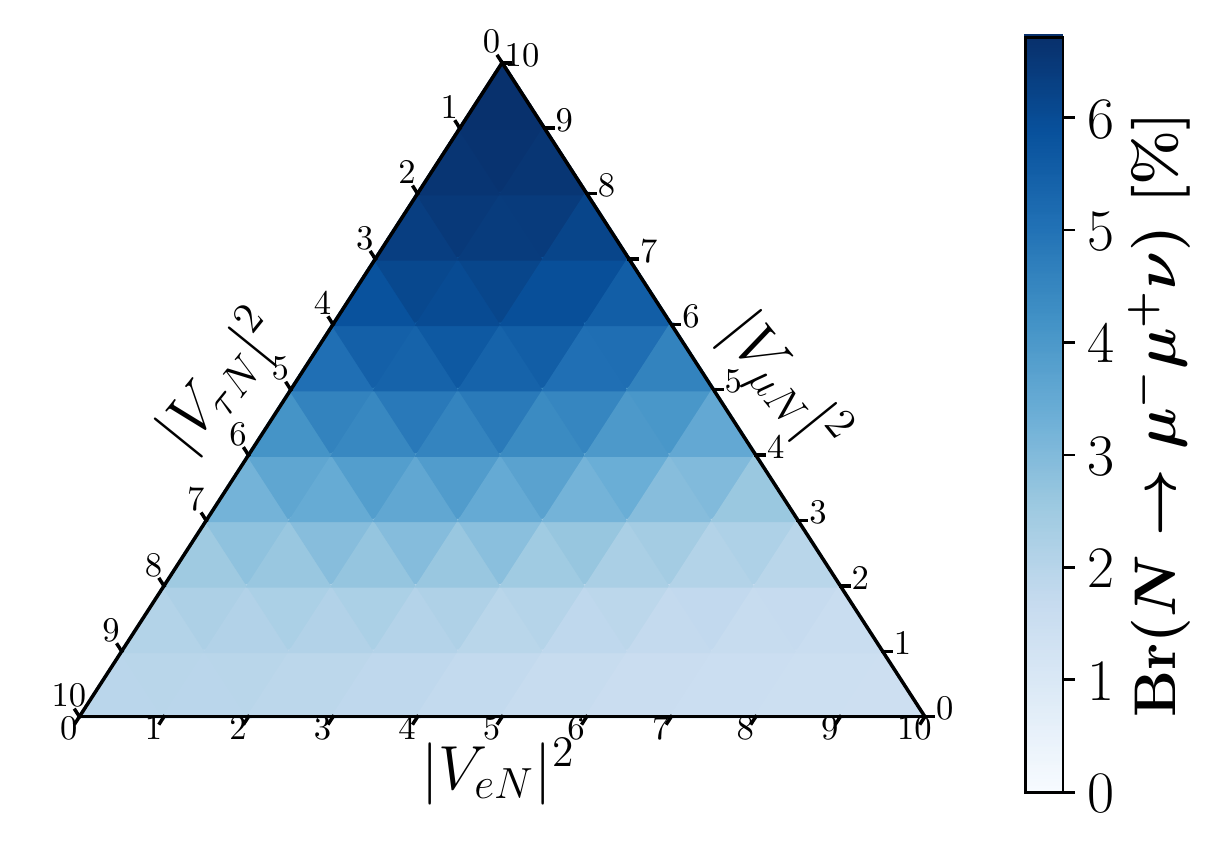}
        \includegraphics[width=.32\textwidth]{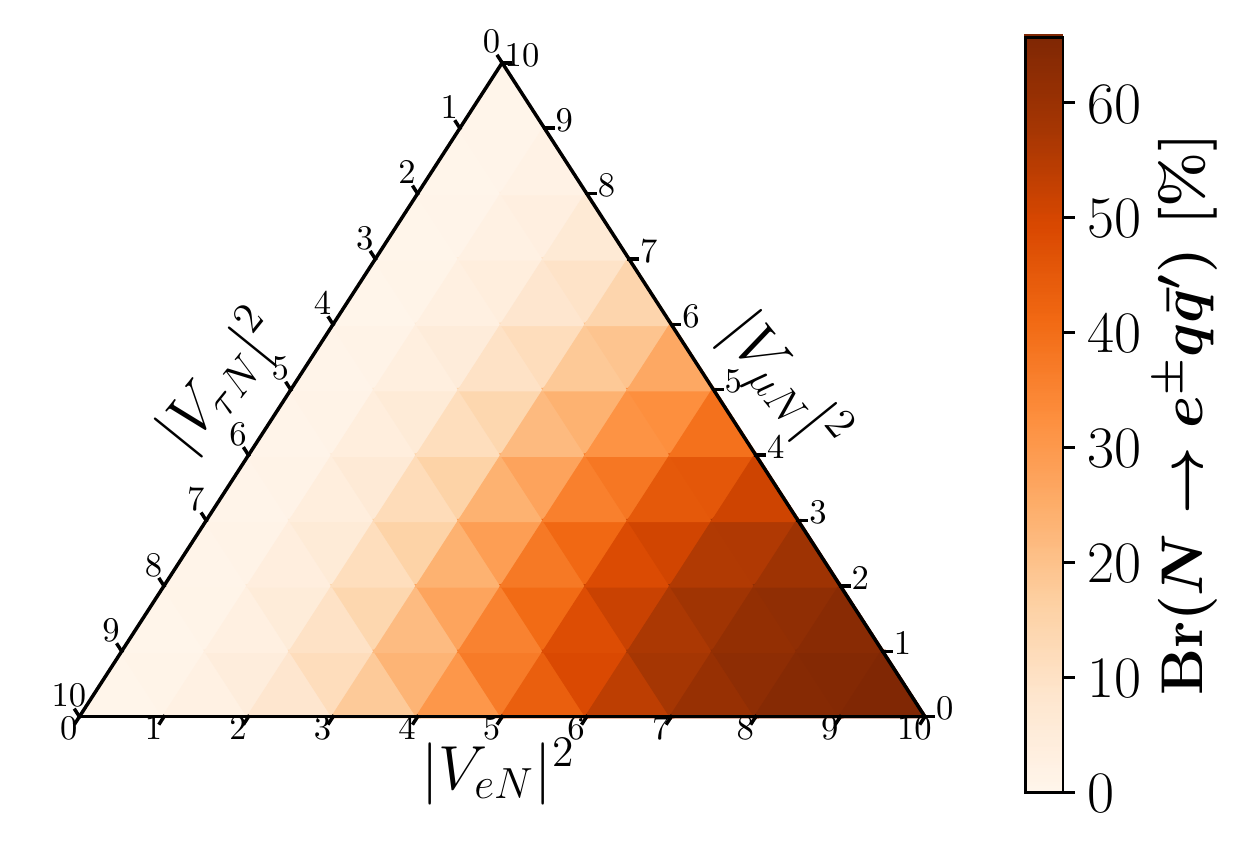}
        \includegraphics[width=.32\textwidth]{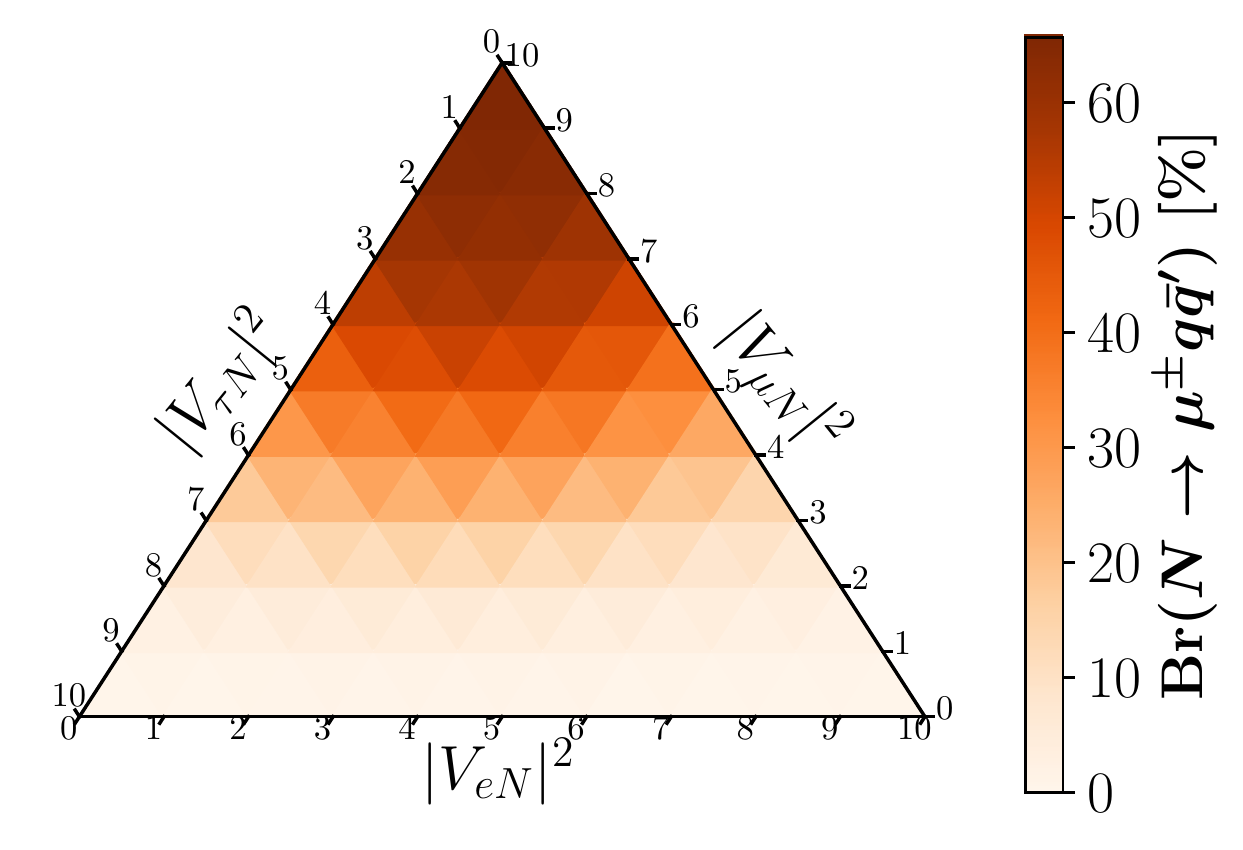}
        \includegraphics[width=.32\textwidth]{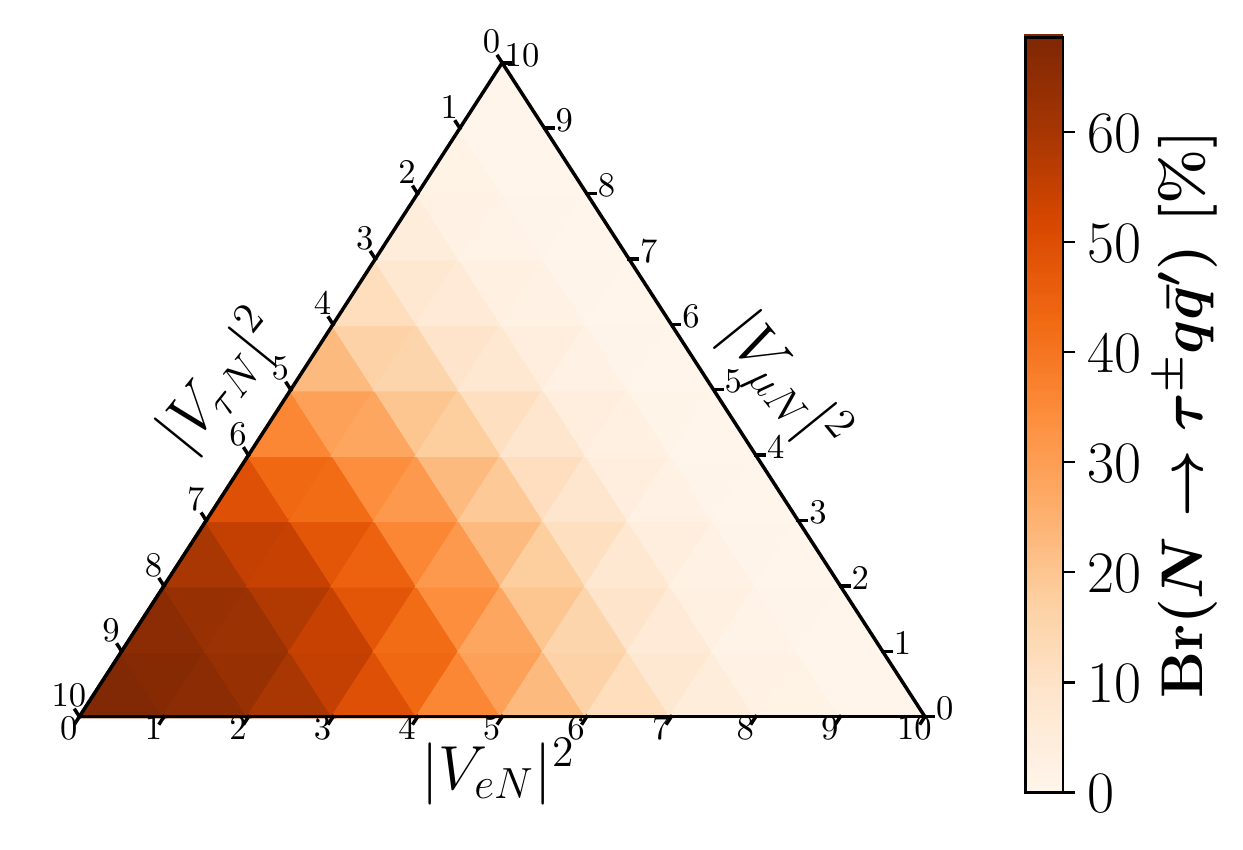}
        \caption{Branching ratios for different decay channels for $m_N=10$~GeV as a function of $|V_{eN}|^2$:$|V_{\mu N}|^2$:$|V_{\tau N}|^2$, since the absolute size of $|V_{\ell N}|^2$ cancels in the ratio. The colors show the different scales chosen to better illustrate the impact of each channel.
		}\label{Decay_BR}
        \end{center}
\end{figure}

Before concluding this section, we focus on the branching ratios for the different HNL decay channels, which are important to understand the kind of signature we can expect in the DV. 
As we already said, in the few GeV mass region that we are interested in, the main decay channels are $N\to \ell\ell'\nu$, $N\to\nu\nu\nu$ and $N\to\ell q\bar q'$. 
Since all of them are three body decays, they will have the same dependence with $m_N$, implying that the branching ratios will be almost mass independent. 
The most relevant dependence will  be that from the mixings $|V_{\ell N}|$, or more precisely from the ratio of mixings $|V_{eN}|^2$:$|V_{\mu N}|^2$:$|V_{\tau N}|^2$, as the total width normalizes their global size. 
In this situation, one can display the different decay channels in triangular plots, like those given\footnote{We do not display the invisible $N\to\nu\nu\nu$ decay as it would be approximately the same in all the triangle, with a value for the BR around  $6\%$.} in Fig.~\ref{Decay_BR}.

The largest branching ratio is for the semileptonic $N\to\ell q\bar q'$ decay, with a total of  approximately $60\%$ divided between the three flavors of the lepton. 
The concrete value of each channel depends on $|V_{eN}|^2$:$|V_{\mu N}|^2$:$|V_{\tau N}|^2$, and therefore the ratio between different final state flavors would provide information on the mixing ratios.
Unfortunately, these semileptonic channels are more difficult to explore in a hadronic collider. 
On the other hand,  purely leptonic channels have a clearer signature, however their branching ratios are smaller, especially those with  leptons of the same flavor.
Interestingly, channels such as $N\to e^\pm\mu^\mp\nu$ have a larger ratio and could be promising to explore, since one could take advantage of having two leptons of a different flavor from a DV. 
We will discuss further the potential of each decay channel is Section~\ref{sec:inclusive}.


\section{Displaced Vertices from Inclusive HNL Production at the LHC}\label{sec:inclusive}

In this work we  consider the production of a HNL in the regime in which there could be a detectable DV, i.e. $m_N$ in the GeV ballpark.
We restrict our analysis to the case in which the displaced decay occurs  in the inner tracker of ATLAS or CMS, although it is possible that the HNL decay occurs in the calorimeters.

The dominant HNL production channels involve mesons for low HNL masses, below the $b$-quark mass~\cite{Bondarenko:2018ptm}, however in this case $i)$ the decay products of such a light HNL are typically very soft and difficult to study at the LHC, and  $ii)$ the HNL tends to be long-lived and to decay outside the detector. 
These production channels are important for beam-dumped experiments such as SHiP~\cite{Alekhin:2015byh,Anelli:2015pba} or for proposed future detectors dedicated to long-lived particles searches~\cite{Kling:2018wct,Helo:2018qej,Curtin:2018mvb}, among others.

\begin{figure}[t!]
\begin{center}
\includegraphics[width=0.49\textwidth]{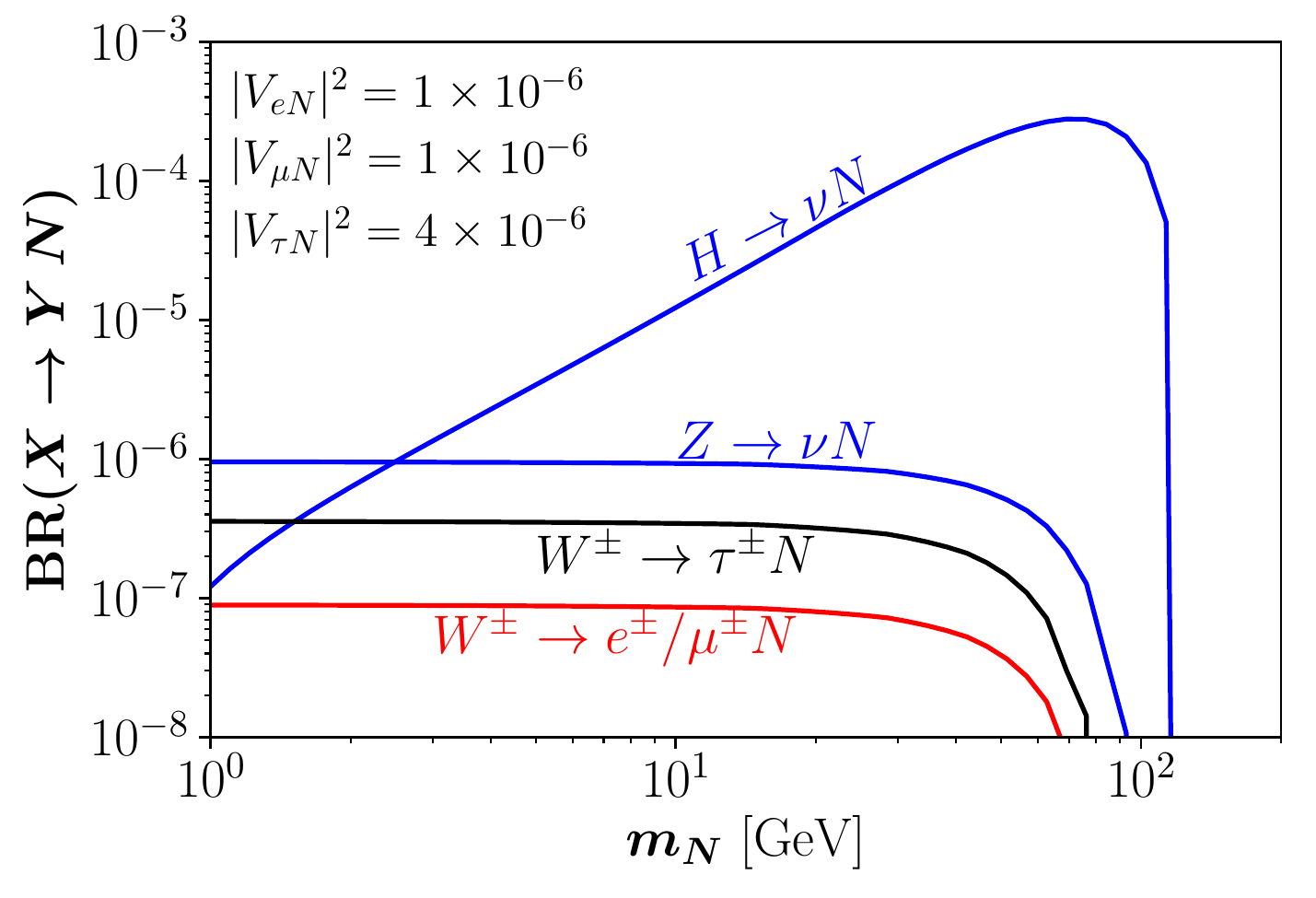}
\includegraphics[width=0.49\textwidth]{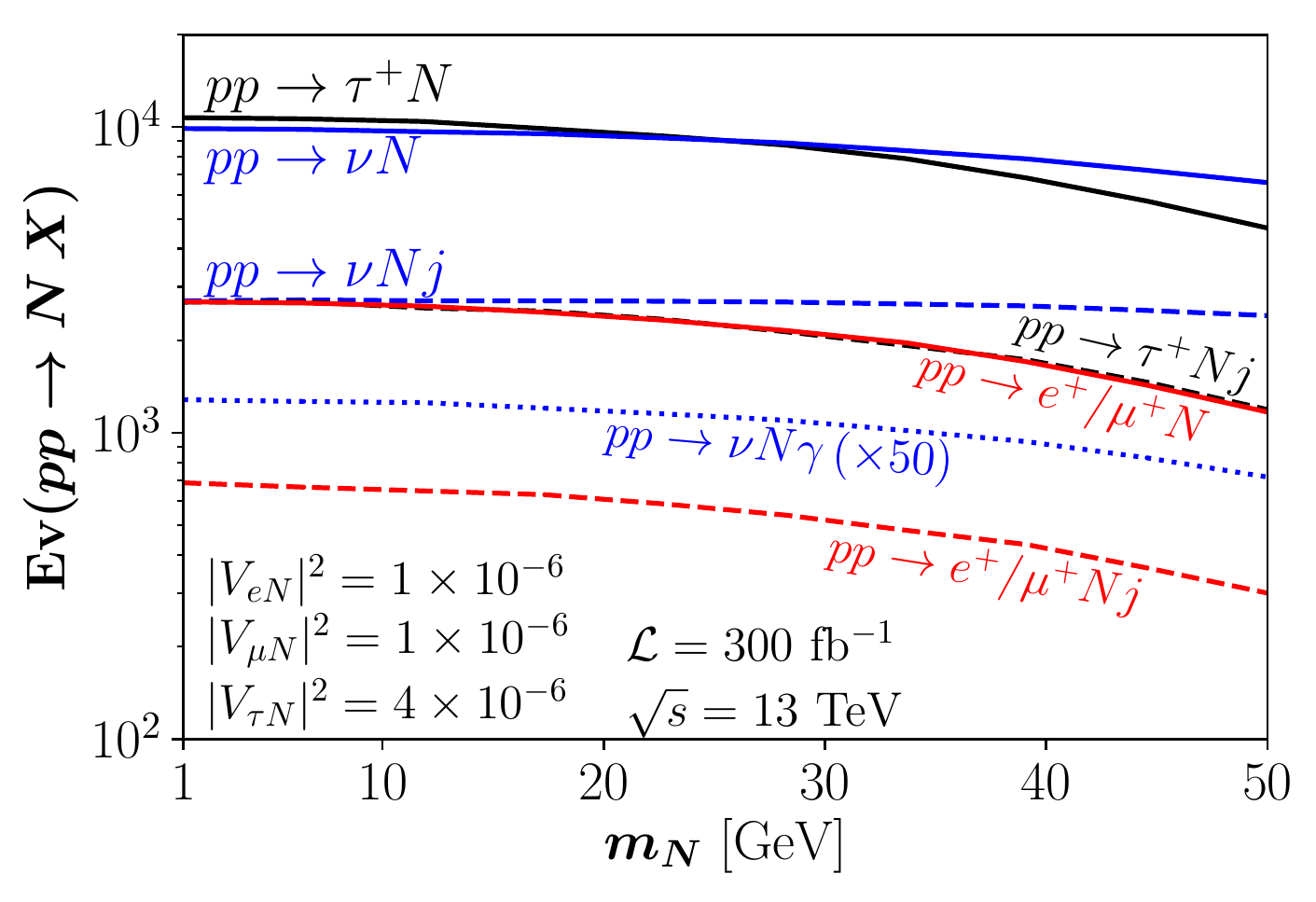}
\caption{Left: Branching ratios for $W^\pm\to\ell^\pm N$, $Z\to\nu N$ and $H\to\nu N$, where $\nu$ stands for the sum over the three light neutrinos.
Right: HNL production at LHC in association with a charged lepton or a light neutrino. We also show the production with an additional jet or photon with $p_T^j>20$~GeV and $p_T^\gamma>10$~GeV, respectively. 
We have chosen $|V_{eN}|^2=|V_{\mu N}|^2=10^{-6}$, $|V_{\tau N}|^2=4\times 10^{-6}$, $\mathcal L=300~{\rm fb}^{-1}$ and $\sqrt{s}=13$~TeV.
}\label{Production-mN}
\end{center}
\end{figure}

For the mass region we are interested in, the relevant production mechanism is via $W^{\pm}$, $Z$ and $H$ bosons that decay into a HNL.
 Some other (subdominant) channels including additional particles are also interesting to explore, since their different kinematics gives access to alternative regions of the parameter space via DV signatures, as will be discussed in Section~\ref{sec:boosted}.
Consequently, we focus on the production from $W/Z/H$ boson decays, which are indeed the main channels for the mass region we are interested in.
Some important features should be noted:
\begin{itemize}
\item The decays from $W^\pm$ and $Z$ bosons are largely independent of $m_N$, for $m_N\ll m_W$ and $m_Z$.
\item The decays from $W^\pm$ bosons are flavor dependent, proportional to a single flavor mixing.
\item The decays from $Z$ and $H$ bosons are flavor independent, proportional to the sum of the square of all mixings, which is the same combination that enters in the HNL total decay  width, Eq.~\eqref{eq:Nwidth}. 
\item The Higgs boson decay has an additional suppression proportional to the square of the ratio of $m_N/m_W$.
\end{itemize}
The branching ratios can then be expressed as:
\begin{equation}
{\rm BR}_{W^\pm\to \ell^\pm N}\propto \big| V_{\ell N}\big|^2\,,~~
{\rm BR}_{Z\to\nu N}\propto \sum_{\ell} \big| V_{\ell N}\big|^2\,,~~
{\rm BR}_{H \to\nu N}\propto \left(\frac{m_N}{m_W}\right)^2\,\sum_{\ell} \big| V_{\ell N}\big|^2\,,
\end{equation}
where for $Z$, $H$ boson decays we have summed over the three light neutrinos, as one cannot distinguish them at the LHC.
Though the contribution from Higgs bosons is subdominant for small values of the HNL mass, this production channel is relevant to infer information on the neutrino  mass generation mechanism.
We show in Fig.~\ref{Production-mN} (left) the explicit values of these branching ratios as a function of HNL mass for an example of the mixing pattern: $|V_{eN}|^2=|V_{\mu N}|^2=10^{-6}$ and $|V_{\tau N}|^2=4\times 10^{-6}$.
Full analytical expressions for these decays can be found e.g. in Refs.~\cite{Abada:2013aba,Gago:2015vma}, and are therefore not reported here.

As a large number of $W^\pm$, $Z$ and $H$ bosons are produced at the LHC, we can set parameter space sensitivity limits for this inclusive DV analysis. We show in Fig.~\ref{Production-mN} (right) the number of HNL produced in $pp$ collisions at $\sqrt{s}=13$~TeV, with an integrated luminosity $\mathcal L=300~{\rm fb}^{-1}$.
We have generated  the $pp\to\ell N$ and $pp\to\nu N$ processes and we have checked that they are indeed dominated by the production and decay of on-shell $W^\pm$ and $Z$ bosons, respectively. We emphasize however that the relative importance of each $W^\pm$ channel depends on the relative size of each mixing, unlike the case of the $Z$ boson decays.
In Fig.~\ref{Production-mN} (right) we show as well the events corresponding to the (also) flavor blind production channels  $pp\to\nu N j$ and $p p\to \nu N \gamma$.
These channels could be useful as the initial state radiated jet or photon could be used as trigger for the interaction point in the $Z$ channel, or to access different kinematic regimes of the HNL. 

In order to estimate the impact of the chosen HNL flavor hypothesis on DV searches,  we define three illustrative  benchmark scenarios for the forthcoming analysis:
\begin{enumerate}
\item Black lines: Mixing to one single flavor, $V_{e N}$, with $V_{\mu N}=V_{\tau N}=0$.
\item Blue lines: Equal mixing to two flavors, $\big|V_{e N}\big|=\big|V_{\mu N}\big|$, with $V_{\tau N} =0$.
\item Green lines: Democratic mixing to three flavors, $\big|V_{e N}\big|=\big|V_{\mu N}\big|=\big|V_{\tau N}\big|$.
\end{enumerate}
Based on our results, we classify the different production channels into flavor dependent and independent categories, showing how the sensitivity predictions of DV searches vary in the former case but not in the latter. 
This motivates us to focus on the inclusive production analysis for the DV with both leptonic and semileptonic HNL decays.


\subsection[Flavor Dependent Production in $pp\to \ell\,N$]{Flavor Dependent Production in $\boldsymbol{pp\to \ell\,N}$}

The details of the leptonic flavor structure are prominent only in this decay channel dominated by $W^\pm$ decays, thus in this case any experimental analysis will be flavor dependent. 
This channel is interesting since the charged lepton, if detected, could be used as a trigger for the primary vertex.
This production mechanism has been explored in the literature~\cite{Helo:2013esa,Izaguirre:2015pga} with the conclusion that the LHC could probe mixings up to $|V_{\ell N}|^2\sim10^{-7}$, with $\ell=e$, $\mu$, after collecting $\mathcal L=300~{\rm fb}^{-1}$ of data,  implying that the LHC could explore this mass range beyond present constraints. In all our figures experimental bounds correspond to the shaded green areas of the plots.
Nevertheless, these analyses assumed that the HNL mixes only to one flavor, which is a condition that is typically not realized in most of the BSM models, and the deviations from this simplified  hypothesis may change the conclusions.

\begin{figure}[t!]
\begin{center}
\includegraphics[width=0.49\textwidth]{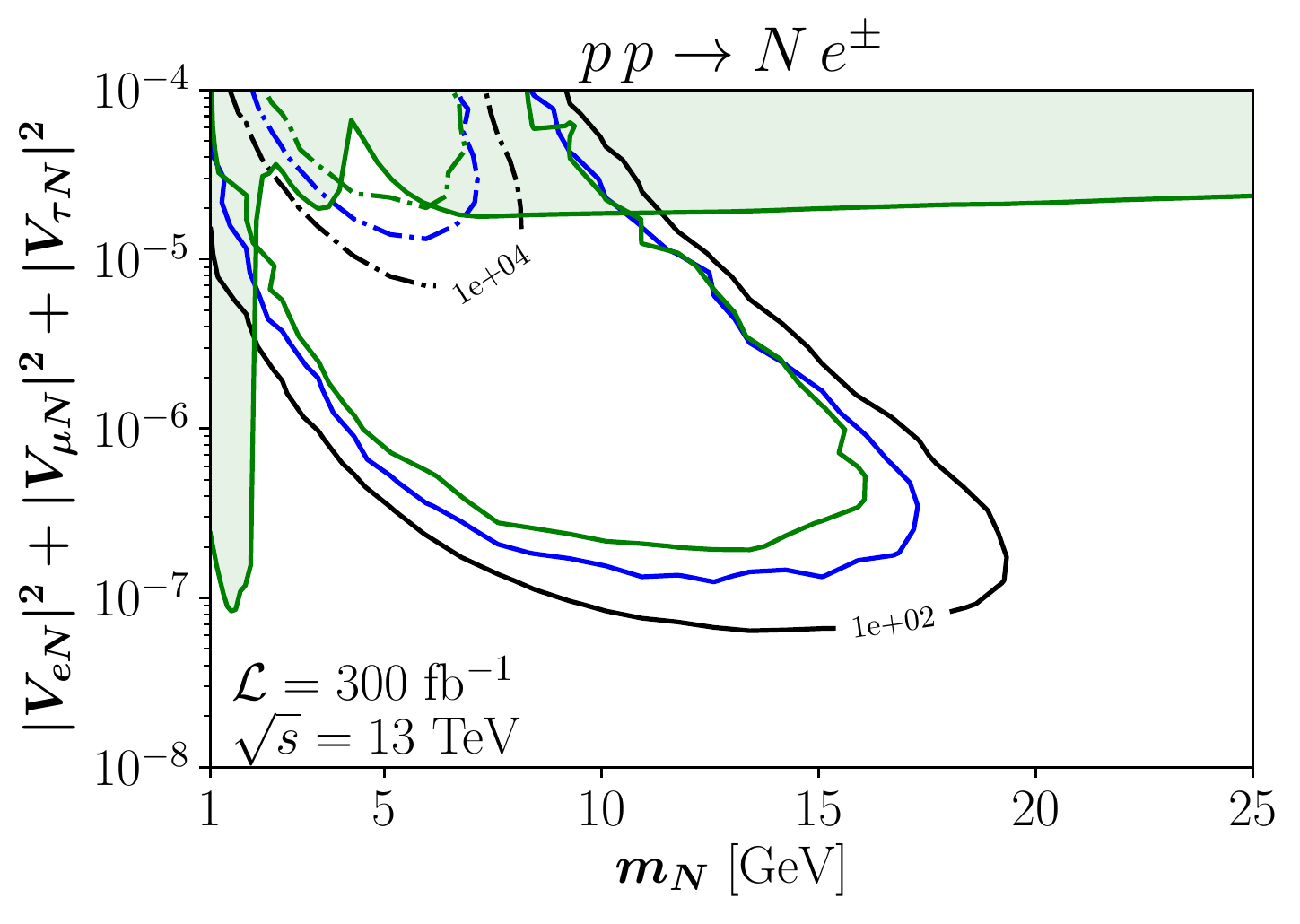}
\includegraphics[width=0.49\textwidth]{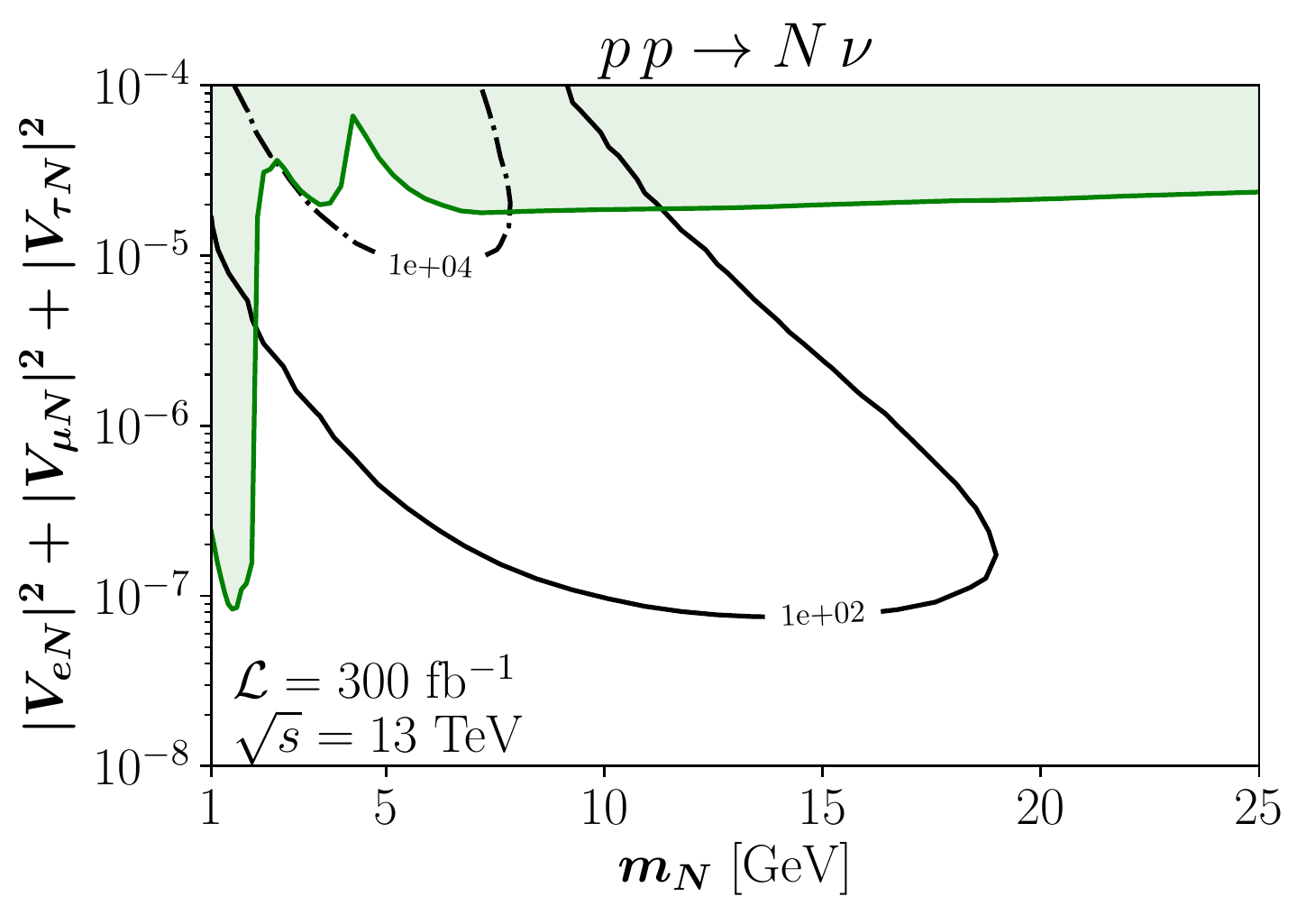}
\caption{Left: HNL production in association with a $e^\pm$, for three scenarios: only one mixing $|V_{eN}|$ (black), two mixings $|V_{e N}|=|V_{\mu N}|$ (blue) and three mixings $|V_{e N}|=|V_{\mu N}|=|V_{\tau N}|$ (green).
Right: HNL production in association with a light active neutrino. Here the three scenarios give the same results.
In both panels, solid (dot-dashed) lines are contour lines for $10^2$ ($10^4$) events with DV at the LHC 13~TeV with $\mathcal L = 300~{\rm fb}^{-1}$.
Cuts on the $e^\pm$ ($p_T^e>25$~GeV and $|\eta^e|<2.5$) and on the DV are imposed.
Shaded green areas are excluded by experimental bounds.
}\label{DVevents-mNVlN}
\end{center}
\end{figure}

The impact of the mixing to different flavors is exemplified in the left panel of Fig.~\ref{DVevents-mNVlN}, where we consider DV events from the HNL production in association with an electron or a positron for the three benchmark scenarios described in the previous section.
We have imposed cuts on the electron and positron ($p_T^e>25$~GeV and $|\eta^e|<2.5$), and on the DV (1~mm $< \ell_{{}_{\rm DV}}< 1$~m and $z_{{}_{\rm DV}}<300$~mm).
The solid (dot-dashed) lines are contour lines for $10^2$ ($10^4$) events.
From this figure we can see that the number of events, and therefore the sensitivity, is different in each scenario, the most optimistic numbers being those of the simplified scenario with only one non-vanishing mixing.
The differences come from the fact that the total width, defining the area where DV may occur, depends on the sum of all mixings, see Eq.~(\ref{eq:Nwidth}), while the production rate of $p p\to \ell N$ is only sensitive to $|V_{\ell N}|$.
This is the same effect we already discussed in Fig.~\ref{DVdistributionsVtauN}.
Therefore, when considering mixings also to other flavors, the relative importance of $|V_{eN}|$ decreases and so does the sensitivity via the $p p\to e^\pm N$ channel.
Moreover, notice that the number of events in Fig.~\ref{DVevents-mNVlN} corresponds to the total number of produced HNL that would decay, to any channel, at a DV. Therefore the full process including the HNL decay to a given channel will have an even stronger dependence on the flavor hypothesis.

\subsection[Flavor Independent Production in $pp\to \nu\,N$]{Flavor Independent Production in $\boldsymbol{pp\to \nu\,N}$}
We have performed  the same exercise for the $p p\to \nu\,N$ production channel, where $\nu$ stands for the sum over the three light neutrinos. The three considered scenarios lead to the same number of DV events, with small differences at low values of $m_N\lesssim m_\tau$, and are thus shown as a single color in the right panel  of Fig.~\ref{DVevents-mNVlN}.
As we explained before, both the total width and the production rate depend on the same combination $|V_{eN}|^2+ |V_{\mu N}|^2+ |V_{\tau N}|^2$.
Therefore, this flavor blind production mechanism is closely related to DV searches, as it was already pointed out when exploring HNL via $H$ decays~\cite{Gago:2015vma}.

The idea of the flavor independent production can be extended to an inclusive search, where we focus only in detecting the DV.
In this case, the production cross section and the decay width have the same flavor dependence, and thus one can set bounds without making further assumptions. 
Of course, a flavor dependence will enter if the HNL decays to charged leptons, nevertheless it can be explored by searching for different final state channels, as discussed in the next section.

\subsection{Inclusive HNL Production and Displaced Vertices}\label{inclusive}

In order to focus on an inclusive HNL production, we consider all possible contributions $pp\to XN$ (referring to  $pp\to\nu\,N$ and $pp\to\ell\,N$ with $\ell=e$, $\mu$, $\tau$) followed by the decay of HNL into different channels, as discussed in Section~\ref{sec:HNLdecay}. 
This inclusive production is flavor blind and it depends on the sum of squared mixings, similar to  the HNL total width. 
Therefore, we show the sensitivity plots in terms of the combination $|V_{e N}|^2+|V_{\mu N}|^2+|V_{\tau N}|^2$, the proper variable to be explored in DV analysis.
We do not place any cuts on the primary vertex decay products. 
However, in order to identify the DV and select the events, we have imposed the following cuts:
\begin{itemize}
	\item 1 mm $< \ell_{{}_{\rm DV}}< 1$~m and $z_{{}_{\rm DV}}<300$~mm, 
	\item $|\eta^\ell| < 2.5$ and $|\eta^j|<2.5$ for tracks from the DV,
	\item $p_T^e>10$~GeV, $p_T^\mu>5$~GeV and $p_T^j>10$~GeV for tracks from the DV,
	\item $m_{\ell\ell '}$, $m_{\ell j j '}<m_N$ when there are two charged leptons or one charged lepton and two quarks from the DV, 
	\item $\Delta R_{\ell \ell'}$, $\Delta R_{j j'}<1$ when there are two charged leptons/quarks from the DV.
\end{itemize}

\begin{figure}[t!]
\begin{center}
\includegraphics[width=0.49\textwidth]{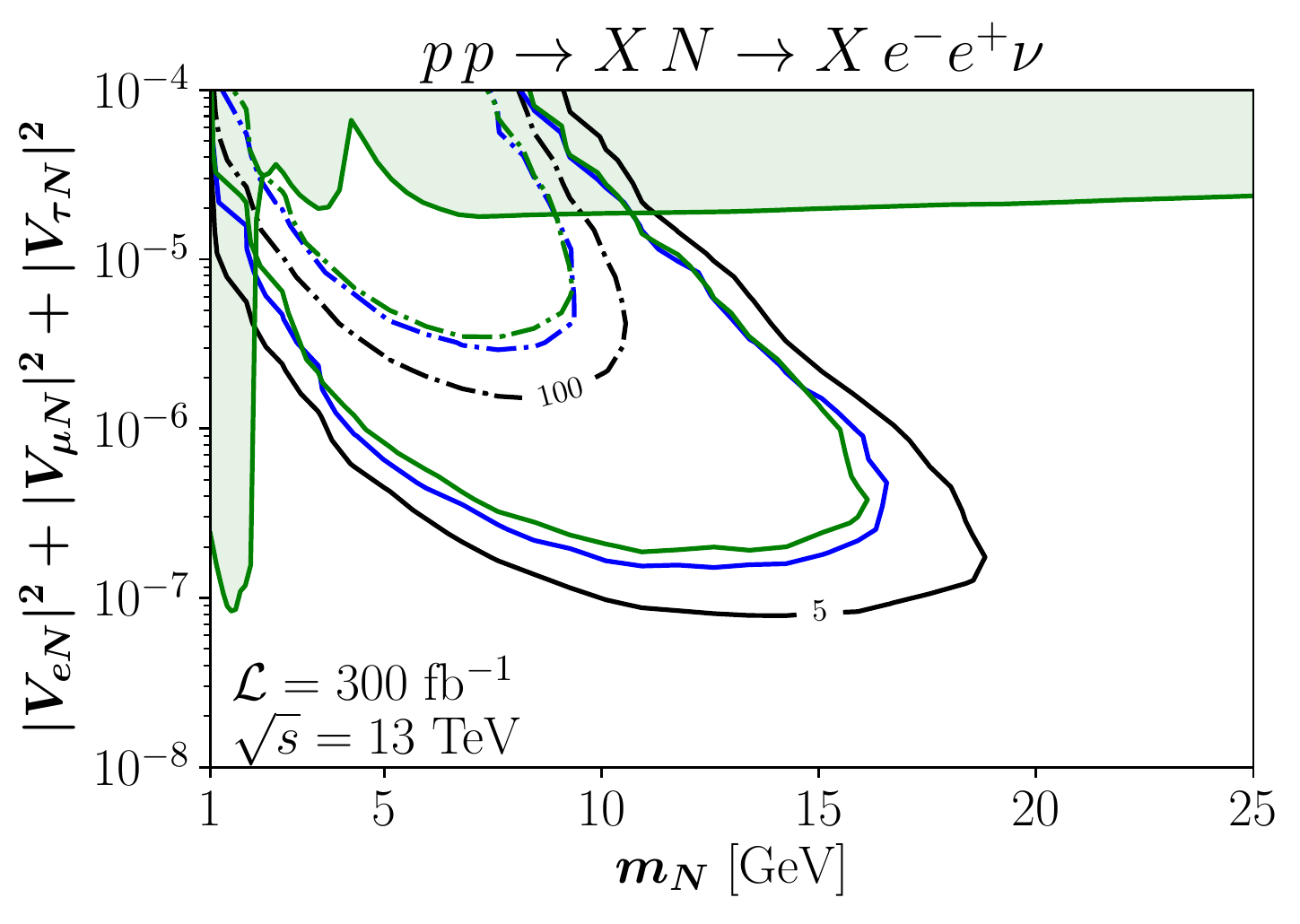}
\includegraphics[width=0.49\textwidth]{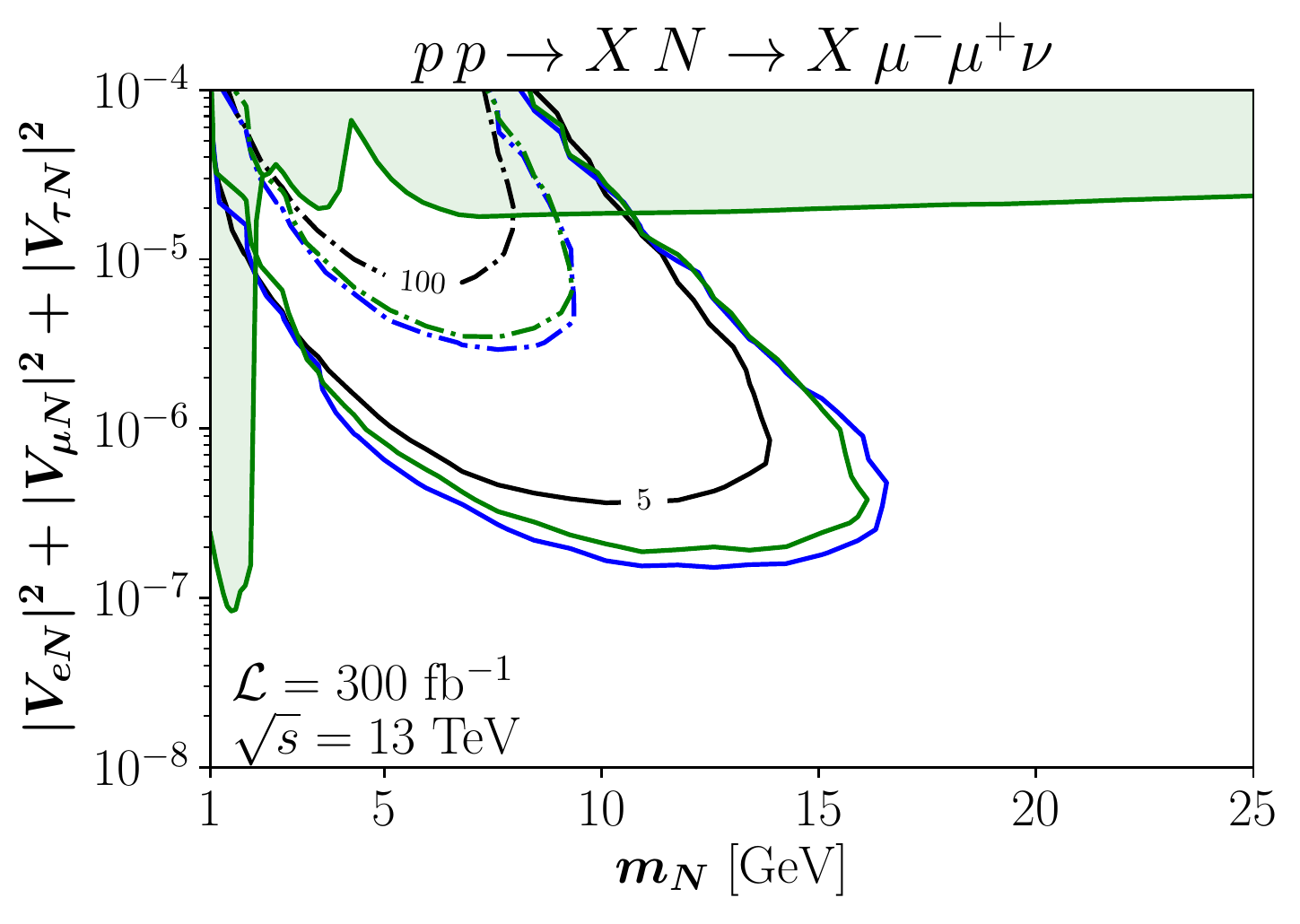}
\includegraphics[width=0.49\textwidth]{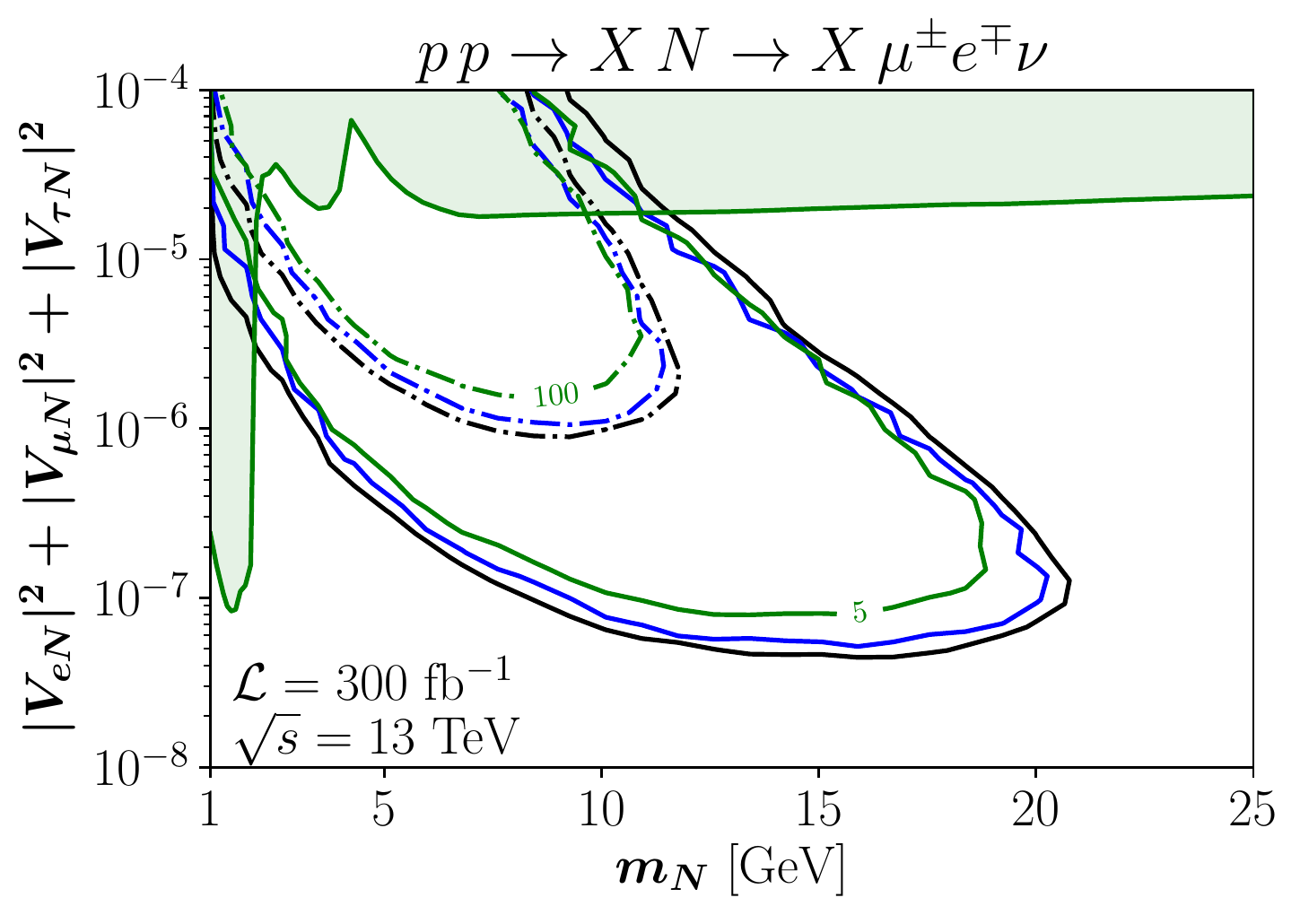}\\
\includegraphics[width=0.49\textwidth]{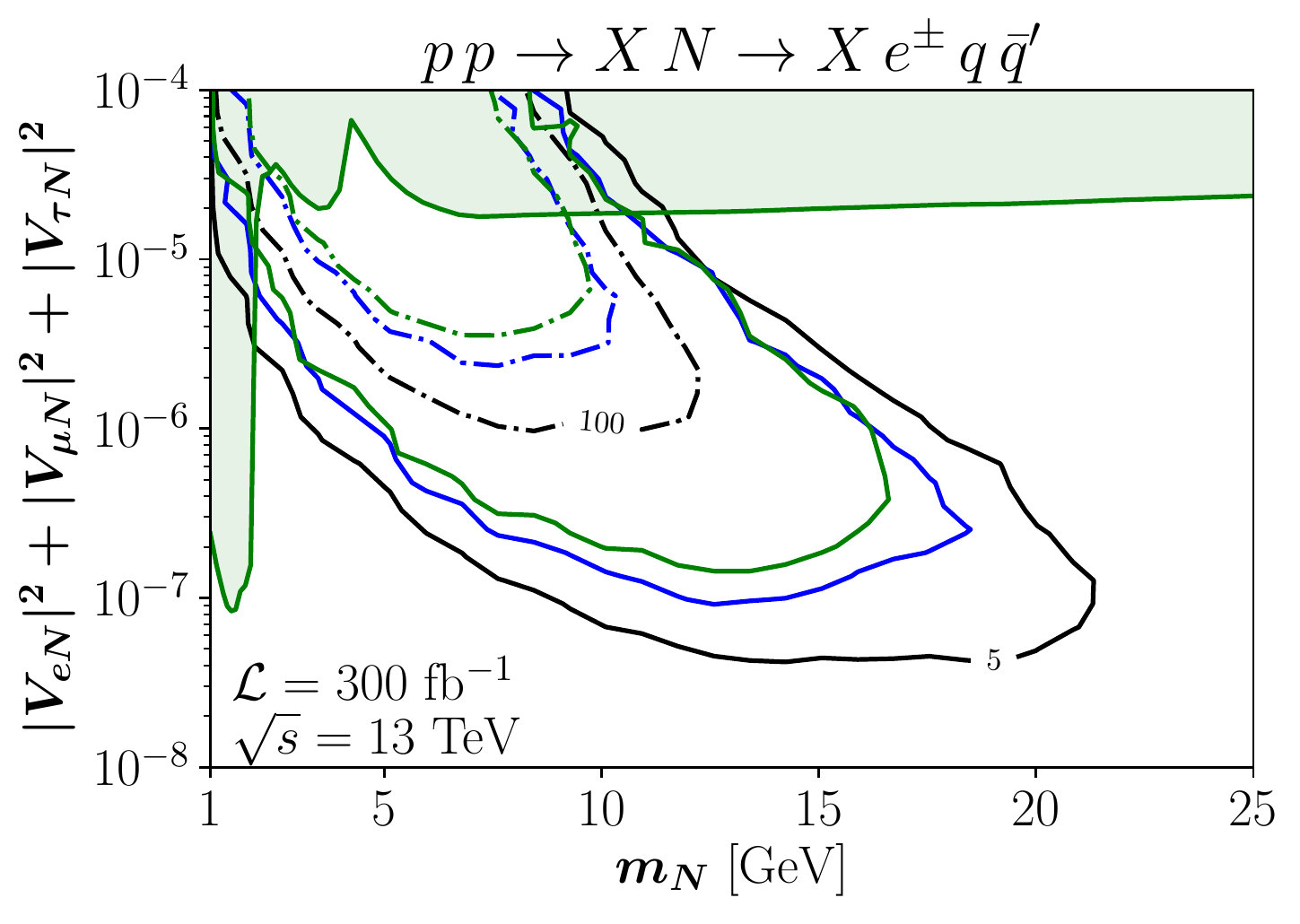}
\includegraphics[width=0.49\textwidth]{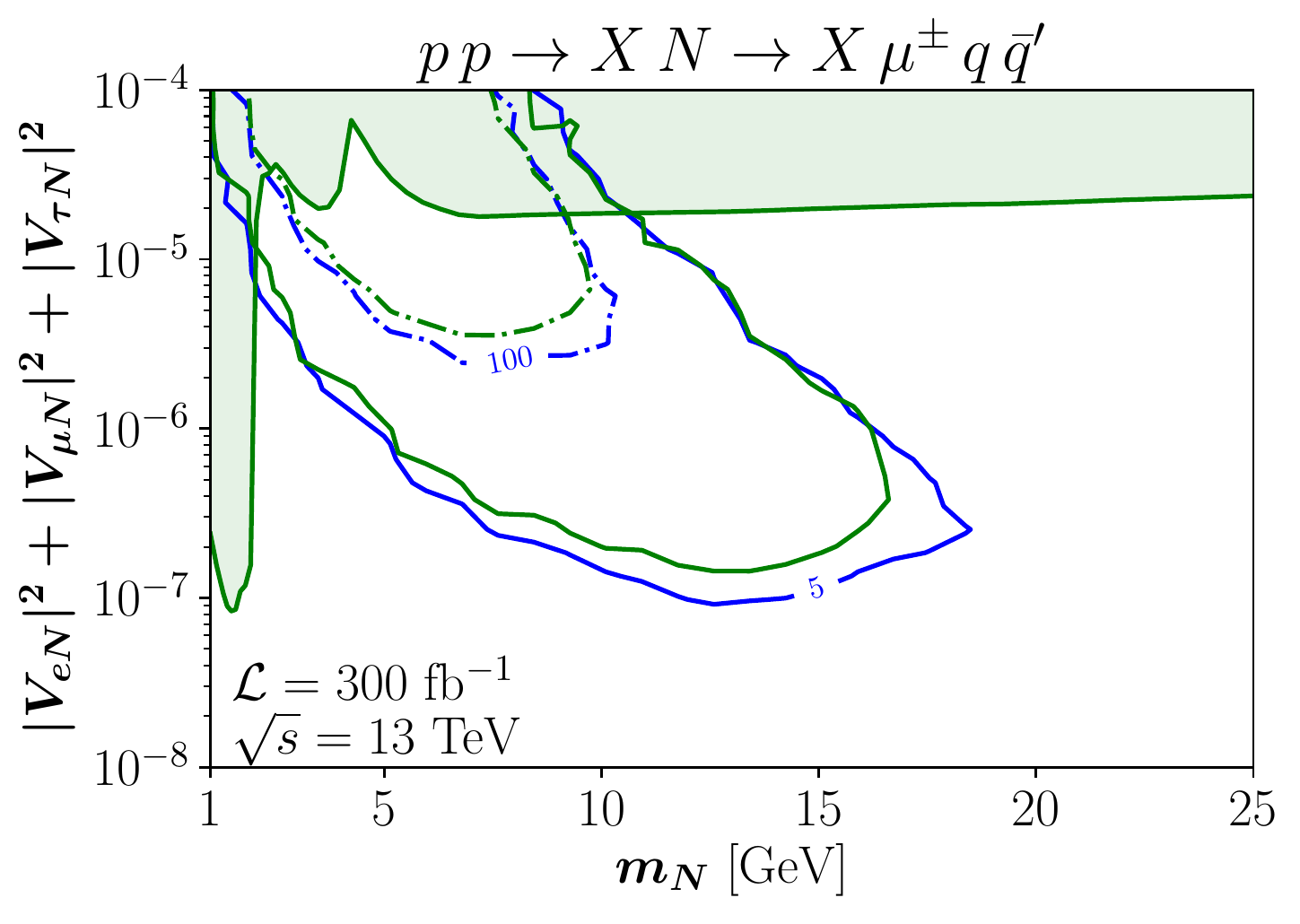}
\caption{Inclusive HNL production followed by its decay into different channels, for mixings $|V_{e N}|^2:|V_{\mu N}|^2:|V_{\tau N}|^2$=1:0:0 (black), 1:1:0 (blue) and 1:1:1 (green).
The solid (dot-dashed) lines are contour lines for $5$ ($100$) events with DV at the LHC 13 TeV with $\mathcal L = 300~{\rm fb}^{-1}$ and the cuts described in the text.
Shaded green areas are excluded by experimental bounds.
}\label{DVevents-inclusive}
\end{center}
\end{figure}

Fig.~\ref{DVevents-inclusive} shows the number of events for the inclusive HNL production and different decay channels: $N\to e^-e^+\nu$, $\mu^-\mu^+\nu$, $\mu^\pm e^\mp$, $e^\pm q\bar q'$ and $\mu^\pm q\bar q'$.
Different flavor hypothesis are explored in different colors, $|V_{e N}|^2:|V_{\mu N}|^2:|V_{\tau N}|^2$=1:0:0 (black), 1:1:0 (blue) and 1:1:1 (green).
The solid (dot-dashed) lines are contours for $5$ ($100$) events with a DV at the LHC 13~TeV with $\mathcal L = 300~{\rm fb}^{-1}$.
The shaded green areas are the experimental bounds.\footnote{Notice that the strong bounds for $m_N\sim 1-2$ GeV are actually valid only for $V_{eN}$ and $V_{\mu N}$, see Fig.~\ref{Decay_mN}. For higher HNL masses, the bounds are directly constraining the sum of the mixings $\sum_\ell |V_{\ell N}|^2$.}

Let us first note that contours in Fig.~\ref{DVevents-inclusive} follow the ones of Fig.~\ref{DVevents-mNVlN} but with a smaller number of events, due to the suppression of the branching fractions and to the selection cuts.
For the channel $pp\to X\,N\to X\,e^-e^+\nu$ (upper left panel of Fig.~\ref{DVevents-inclusive}), the maximum sensitivity is reached in the case where the HNL mixes only to electrons (black lines).
However, if the HNL is allowed to mix with the muons (blue lines) or the muons and the taus (green lines), other decay channels open up, reducing the branching ratio of the HNL into $e^-e^+\nu$ and therefore decreasing the sensitivity to that particular channel.
A similar behavior is featured in the $pp\to X\,\mu^\pm e^\mp\nu$ channel (central panel), although the flavor dependence is milder.
In this case there are more events due to the larger branching ratios to this channel, and the experimental signature is more interesting because of the different flavor of the final charged leptons.
On the contrary, for the $pp\to X\,\mu^-\mu^+\nu$ channel (upper right panel), the sensitivity is minimal (but not zero) for the case when $V_{\mu N}=0$.
In fact, even if the HNL cannot decay via the $W^\pm$ bosons, it can always via the exchange of a $Z$ boson.
The sensitivity to this channel grows with the mixing $|V_{\mu N}|$ and, in this sense, it is complementary to the purely electronic channel.
In the forthcoming section we will refine the analysis for this channel.

The case for the semileptonic decays (lower panels) resembles the pure leptonic one.
However, some comments are in order.
First, the $pp\to X\,\mu^\pm q\bar q'$ channel (lower right panel) vanishes in the scenario where $V_{\mu N}=0$, because the corresponding decay can only occur by the mediation of a $W^\pm$.
Second, the semileptonic branching ratios are typically larger, and so is the sensitivity.
Finally, the possible backgrounds are also more important than in the full leptonic channels.
Nevertheless, one would like to finally explore all the possible final states, as they provide complementary information about the flavor structure of the HNL.

\subsection{Experimental Sensitivity to the Dimuon Channel}
\label{sec:sensitivities} 

In the previous sections, we have studied the impact of choosing different benchmark scenarios taking into account the flavor dependence in the production when predicting the number of DV events, as well as the importance of exploring all possible HNL decay channels.
Nevertheless, in order to estimate final LHC sensitivities, it is necessary to further discuss event-triggering and potential backgrounds, which have to be studied in detail for each individual decay channel. 
Here, we focus on the dimuon $N\to\mu^+\mu^-\nu$ channel, since muons are particularly interesting in DV searches due to the large volume covered by the muon spectrometers, although equivalent  analysis would be needed for other channels. 

This dimuon channel  with DV signatures has been previously explored in Refs.~\cite{Izaguirre:2015pga,Dube:2017jgo}, considering a lepton jet topology, although working always in the single mixing (flavor) scenario. 
In our analysis, however, we will closely follow the recent ATLAS analysis in Ref.~\cite{Aaboud:2018jbr}, where this kind of dimuon events are triggered by requiring two opposite-sign collimated muons, with transverse momenta larger than 15 and 20~GeV, pseudorapidity $\eta<2.5$ and a small angular separation of $\Delta R_{\mu\mu}<0.5$.
Similar triggers have been also used by CMS, see for instance Ref.~\cite{CMS:2014hka}.

The main SM backgrounds for this dimuon channel may come from low-mass Drell-Yan processes and from single and top pair production.
Other possible sources for backgrounds could be cosmic-ray muons, muons with relatively low momentum in multi-jet events, or muons from long-lived mesons, this latter being important at low dimuon invariant masses.
A full discussion of potential backgrounds can be found in Refs.~\cite{Dube:2017jgo,Aaboud:2018jbr}.

The dimuon invariant mass $m_{\mu\mu}$ is an important variable for this analysis. 
In Ref.~\cite{Aaboud:2018jbr} a lower cut of 15~GeV is imposed.
Nevertheless, we want to emphasize that, in order to explore the light HNL masses leading to DV signatures, one needs to consider very low values for $m_{\mu\mu}$. 
The reason is that the dimuon pair is produced from an on-shell HNL, and thus its invariant mass  will be smaller than the HNL mass.
We will therefore consider that the threshold on $m_{\mu\mu}$ can be lowered down to 5~GeV. 
Even smaller values are experimentally challenging at present LHC detectors. 
Moreover, as we discussed before, HNL masses bellow the bottom mass are dominantly produced in meson decays and, thus, they could be studied in other facilities such as beam-dumped experiments. 
Additionally, we also impose an upper cut of $m_{\mu\mu}<m_N$, reducing potential background from resonant $Z$ boson production and other events leading to high invariant masses.

In order to reduce backgrounds from top production and multi-jet processes, we require muons to be isolated from jets~\cite{Aaboud:2018jbr} and that there is a low hadronic activity~\cite{Dube:2017jgo}.
In particular, we ask for an angular separation between any jet and the muons of 
\begin{equation}\label{cutDeltaRmuj}
\Delta R_{\mu j}> {\rm min}\left(0.4,\,0.04 + \frac{10\,{\rm GeV}}{p_T^\mu}\right)\,,
\end{equation}
and also a track-based isolation criteria of
\begin{equation}\label{cutID}
I^{\rm ID}_{\Delta R=0.4}\equiv\frac{\sum_{ j} |p_T^{ j}|}{p_T^\mu}<0.05\,,
\end{equation}
where the sum goes over all the jets satisfying $p_T^j>0.5$~GeV and $\Delta R_{\mu j} < 0.4$.
High hadronic activity is vetoed by demanding $H_T<60$~GeV, where $H_T$ is the scalar sum of $p_T$ of all  the jets with $p_T>30$~GeV.

Finally, cosmic-ray background is removed by requiring that the two displaced muons satisfy~\cite{Aaboud:2018jbr} 
\begin{equation}\label{cutcosmic}
\sqrt{ \big(\Delta\eta_{\mu\mu}\big)^2 + \big(\pi-\Delta\phi_{\mu\mu}\big)^2} > 0.1\,.
\end{equation}
\begin{table}[t!]
\begin{center}
\begin{tabular}{|c||c|c|}
\hline&&\\[-2ex]
Signal acceptance after cut &   
$\begin{array}{c} m_N=10~{\rm GeV} \\ \Sigma_{\scriptscriptstyle \ell} |V_{\ell N}|^2  = 10^{-6}\end{array} $  &  
$\begin{array}{c} m_N=15~{\rm GeV} \\ \Sigma_{\scriptscriptstyle \ell} |V_{\ell N}|^2  = 10^{-7}\end{array} $  
\\&&\\[-2.2ex]
\hline
\hline&&\\[-2ex]
1~mm $<\ell_{\rm DV}<$ 1~m, $z_{\rm DV}< $ 300~mm & 
66\% & 73\% \\
\hline&&\\[-2ex]
$|\eta_\mu| <$  2.5 &
43\% & 46\%  \\
\hline&&\\[-2ex]
$p_T^{\mu_1} > 20$~GeV, $p_T^{\mu_2} > 15$~GeV & 
1.7\% & 1.4\%  \\
\hline&&\\[-2ex]
$\Delta R_{\mu\mu} < 0.5 $ & 
1.6\% & 1.0\%  \\
\hline&&\\[-2ex]
5 GeV $<m_{\mu\mu} < m_N$ & 
 1\% & 0.9\% \\
\hline&&\\[-2ex]
jet-isolation Eqs.~\eqref{cutDeltaRmuj} and~\eqref{cutID} & 
1\% & 0.9\%  \\
\hline&&\\[-2ex]
$H_T<60$~GeV & 
0.8\% & 0.7\% \\
\hline&&\\[-2ex]
Cosmic veto Eq.~\eqref{cutcosmic} &
0.8\% & 0.7\%  \\
\hline
\end{tabular}
\caption{Cut flow for two HNL benchmark scenarios. 
Each row shows the acceptances with respect to the initial cross sections. We apply the cuts  sequentially from top to down.}\label{cutflow}
\end{center}
\end{table}

\begin{figure}[t!]
\begin{center}
\includegraphics[width=0.49\textwidth]{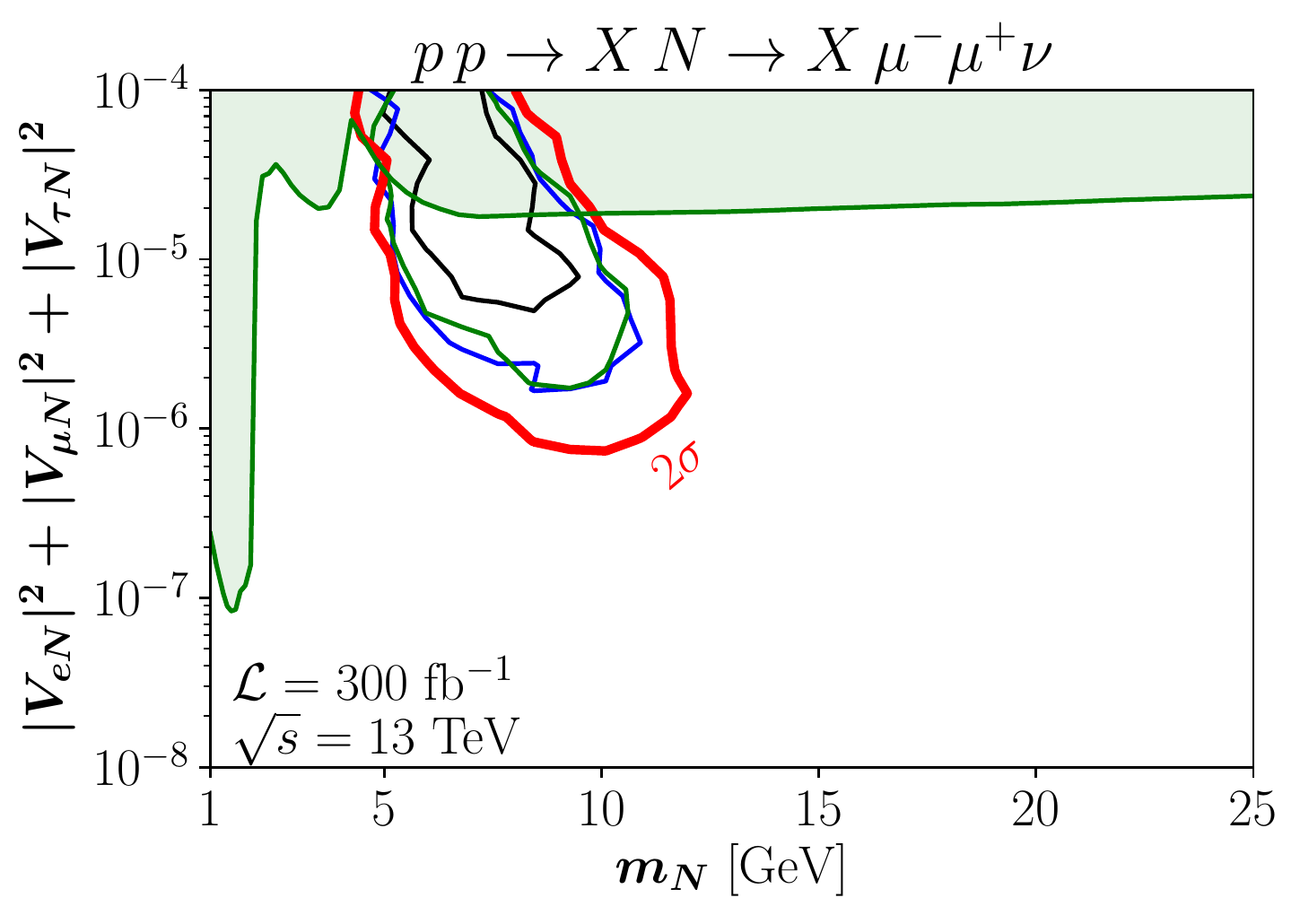}
\includegraphics[width=0.49\textwidth]{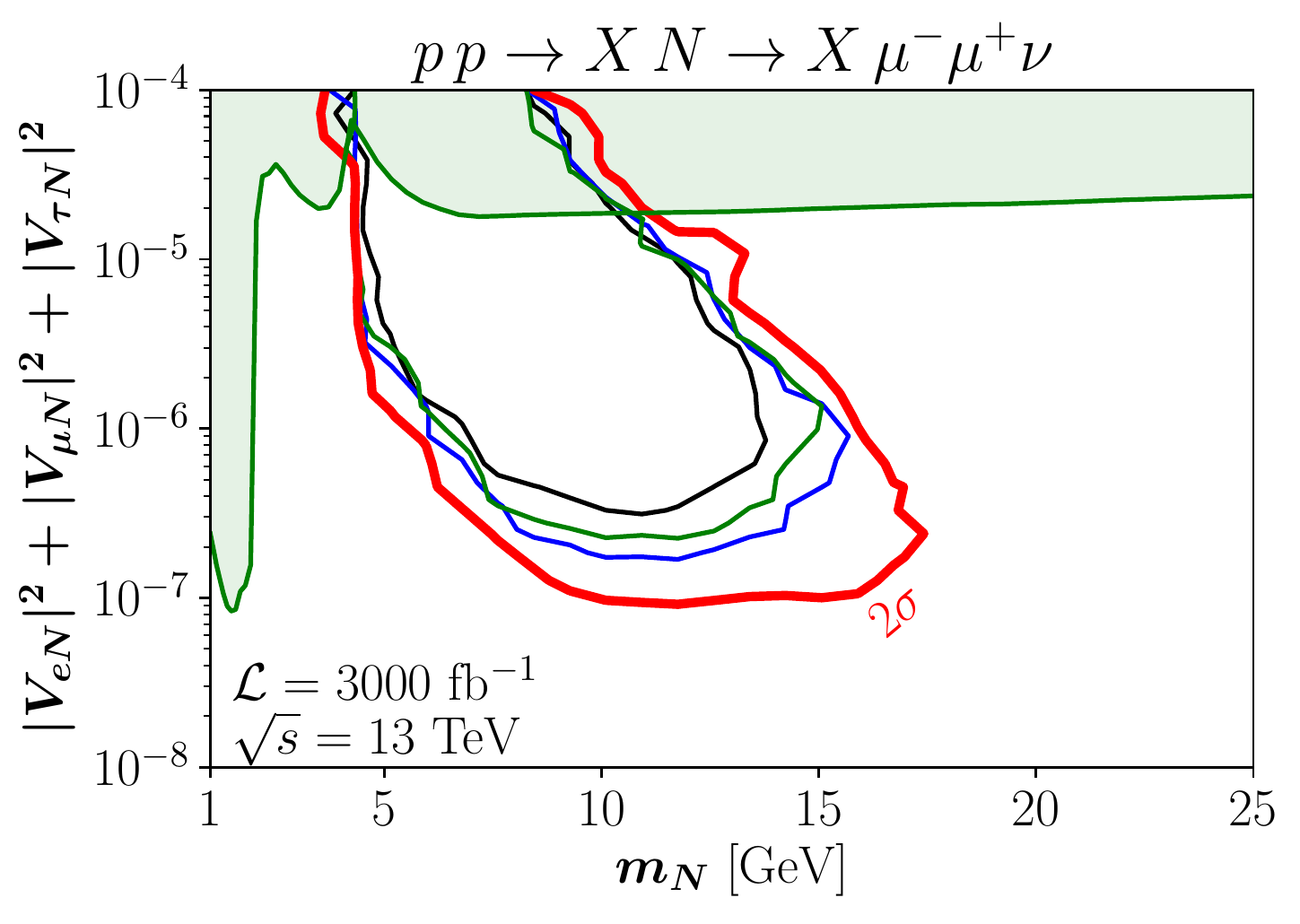}
\caption{Inclusive HNL production followed by its decay into different channels, for mixings $|V_{e N}|^2:|V_{\mu N}|^2:|V_{\tau N}|^2$=1:0:0 (black), 1:1:0 (blue), 1:1:1 (green) and 0:1:0 (thick red).
The solid lines are contour lines for $2\sigma$ exclusion with DV at the LHC 13 TeV with $\mathcal L = 300~{\rm fb}^{-1}$ (left) and $\mathcal L = 3000~{\rm fb}^{-1}$ (right) and the cuts described in section~\ref{sec:sensitivities}.
Shaded green areas are excluded by experimental bounds.
These sensitivity lines are obtained under the background-free hypothesis, see the discussion in the text.
}\label{DVevents-inclusive-ATLAS}
\end{center}
\end{figure}

We list the complete cut flow for two benchmark scenarios in Table~\ref{cutflow}, where we also include the DV condition, as discussed in Section~\ref{inclusive}.
We will assume that these cuts remove all possible backgrounds of DV coming from SM processes.
This optimistic situation is typically adopted in the literature, and is well motivated by the non-observation of any displaced event in analysis such as the one of Ref.~\cite{CMS:2014hka}.
However, this scenario could be not completely realistic. 
Indeed, following the number of events observed in the {\it low-mass} regime of Ref.~\cite{Aaboud:2018jbr}, one could expect a background of the same order of the HNL signal.    
This would deteriorate the sensitivities, pushing the lines in Fig.~\ref{DVevents-inclusive-ATLAS} to  higher values for the couplings, as we will comment later.  
Nevertheless,  the defined signal region differs from ours in both $m_{\mu\mu}$ and the displacement conditions, and therefore it is difficult to conclude on the number of background events for our signal.
In order to obtain the exact effect, a dedicated study with full background treatment is needed, which is largely beyond the scope of this work.

In what follows, we will focus first on the optimistic background-free hypothesis, which is enough to discuss the main ideas in this work. 
Under such hypothesis, 
contours corresponding to the $2\sigma$ exclusion reach can be defined by requiring 3 signal events after cuts.
We show in Fig.~\ref{DVevents-inclusive-ATLAS} the potential LHC sensitivity after collecting $300~{\rm fb}^{-1}$ (left panel) and $3000$~fb$^{-1}$ (right panel) in the background-free hypothesis.
We follow the same color convention as defined previously, and we add an extra thick red line where we assume  that the only non-zero mixing is  $V_{\mu N}$.
This new line helps the comparison with other works in the literature, as well as to better illustrate the effect of changing the HNL flavor hypothesis in the final expected sensitivity areas.
From Fig.~\ref{DVevents-inclusive-ATLAS}, it is clear that the LHC with $\mathcal{L}=300~(3000)~{\rm fb}^{-1}$ luminosity could further probe the parameter space up to one (two) order (orders) of magnitude  below  present bounds.
The exact sensitivity reach in mass and mixings depends however on the chosen HNL flavor pattern and, therefore, we stress again on the importance of exploring all possible final flavor channels.

The previously used background-free hypothesis could be too optimistic, as we said before, since background events could come from dimuon pairs misidentified as displaced events due to detector resolution effects or from random track crossings.
In order to have a rough estimate of how much the inclusion of a background could worsen the sensitivity, we follow the discussion in Ref.~\cite{Deppisch:2018eth}, and assume the largest possible background that is in agreement with not having observed any background event in the CMS analysis of Ref.~\cite{CMS:2014hka}.
This means up to 3 background events for $\mathcal L=20.5~{\rm fb}^{-1}$, which can be scaled to 45 and 450 background events for luminosities of $300~{\rm fb}^{-1}$ and $3000~{\rm fb}^{-1}$, respectively.
Assuming this pessimistic scenario, we have checked that the contours in  Fig.~\ref{DVevents-inclusive-ATLAS} are worsen by approximately one order of magnitude.
This would imply that the LHC could still provide new information about long-lived HNL after collecting $300~{\rm fb}^{-1}$, although the high luminosity phase would be required for exploring all the different flavor patterns shown in Fig.~\ref{DVevents-inclusive-ATLAS}.
Nevertheless, we stress again that a complete analysis of the background is needed in order to have more realistic LHC sensitivities. 

Finally, it is worth  noticing that the sensitivities obtained in Fig.~\ref{DVevents-inclusive-ATLAS} (left) are worst that the ones expected from Fig.~\ref{DVevents-inclusive} (upper-right).
The main reason is the more stringent cuts and trigger requirements, especially those on the lepton transverse momenta, see  Table~\ref{cutflow}.
As discussed before, DV conditions for HNL require that they are light, in the GeV regime, implying soft final leptons that do not pass standard cuts.
This makes these searches more challenging, requiring  cuts and triggers as low as possible in order to increase the signal acceptance.

\begin{figure}[t!]
\begin{center}
\includegraphics[width=0.49\textwidth]{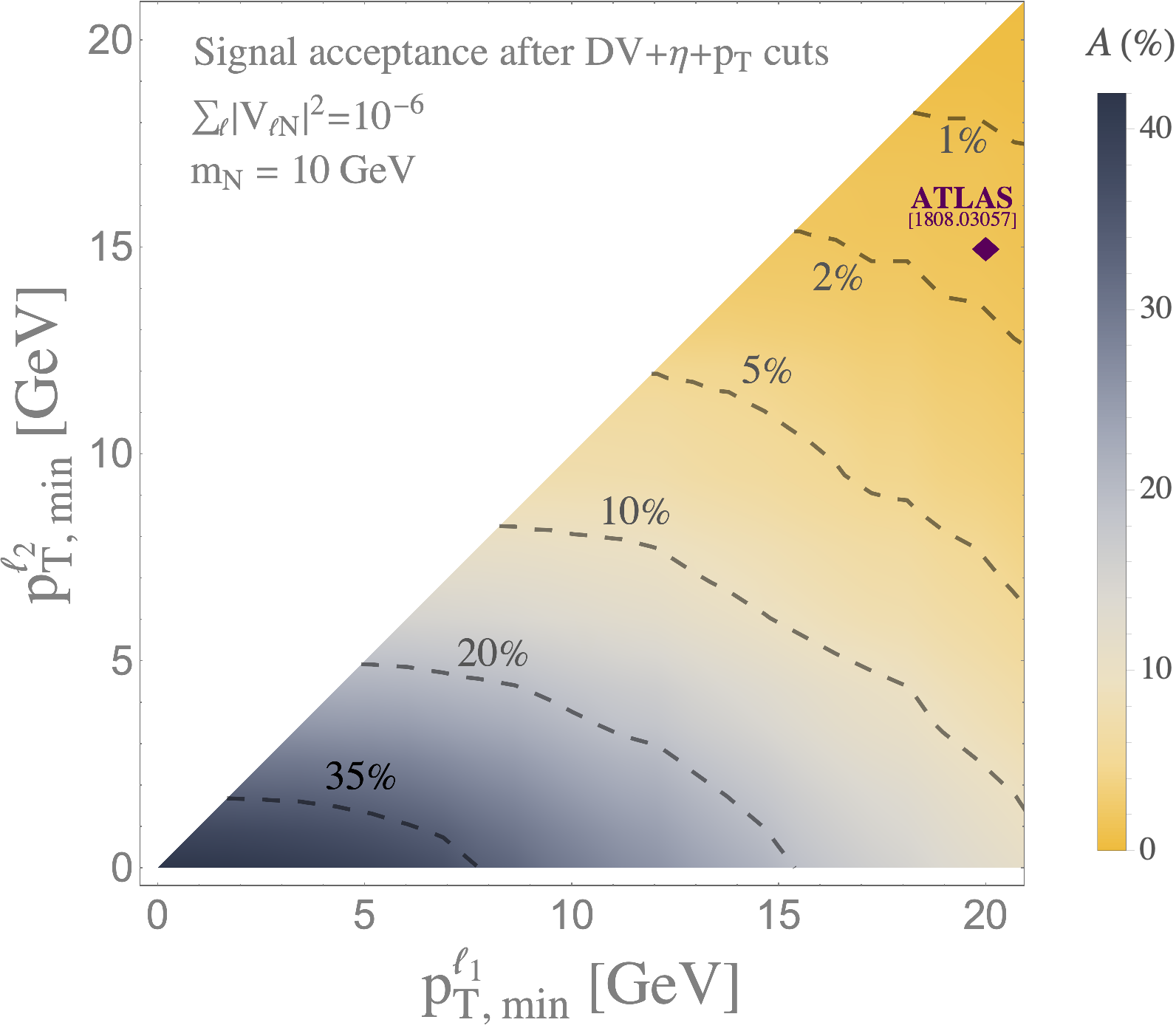}
	\caption{Signal acceptance ($A$) for the $p p \to XN\to X\mu^-\mu^+\nu$ channel after the first three cuts in Table~\ref{cutflow}, varying transverse momenta cuts in the $[0, 20]$ GeV interval. We define $\ell_1$ as the leading lepton, $p_T^{\ell_1}>p_T^{\ell_2}$.
	The diamond marks the cuts we used following ATLAS dimuon analysis~\cite{Aaboud:2018jbr}.
	}\label{NewPTcutsPlot}
\end{center}
\end{figure}

In order to quantify this effect, we display in Fig.~\ref{NewPTcutsPlot} the signal acceptance ($A$) after having applied different cuts on the final lepton transverse momenta, varying them in the $[0,\,20]$~GeV interval; 
in addition,  we applied DV and pseudorapidity constraints for the sake of comparison with Table~\ref{cutflow}. 
One can see that the acceptance of ATLAS triggers in Ref.~\cite{Aaboud:2018jbr} are of about 1-2\%, while the ones proposed in section~\ref{inclusive} are of 20\%, explaining thus the main differences between Figs.~\ref{DVevents-inclusive} and~\ref{DVevents-inclusive-ATLAS}.
Additionally, the signal acceptance can be improved further applying $p_T$ cuts only to one of the muons.
For instance, applying a single cut of 20~GeV the acceptance becomes $\sim 13\%$ and, if that cut is reduced to 5~GeV, the acceptance goes over $35\%$.
Consequently, being able to lower these lepton $p_T$ thresholds or triggering on only one muon would help enhancing sensitivities for displaced HNL decays.

\subsection{Two Heavy Neutral Leptons}
\label{sec2HNL}

We close this section by discussing the possibility of having more than one HNL in our region of interest.
In general, the 3+1 model already captures most of the collider implications of the existence of a  HNL. 
Nevertheless, in seesaw-like models at least two heavy neutrinos are needed in order to accommodate neutrino oscillation data and, consequently, some remarks should be made in this latter case with two HNL, which we effectively describe by a 3+2 model.

Under the lack of a positive experimental signal, one might consider the possibility of reinterpreting the bounds on the 3+1 model in terms of the 3+2 parameters.
The production cross section in the 3+2 would be the sum\footnote{We consider that the two HNL are not degenerate in mass and sum both contributions incoherently. This assumption is justified when the HNL are both long-lived. A more detailed discussion can be found in Ref.~\cite{Chao:2009ef}, where it is stressed  that a  constructive interference of the contributions from two heavy Majorana neutrinos may enhance the cross sections by a factor up to four.}
of the 2 HNL contributions and, if they do not have a large  separation in mass, one can set bounds on the sum $|V_{\ell N_1}|^2 + |V_{\ell N_2}|^2$.
However, the decay length of each HNL depends on each mixing separately, and therefore the interpretations in DV searches are more involved.

\begin{figure}[t!]
\begin{center}
\includegraphics[width=.495\textwidth]{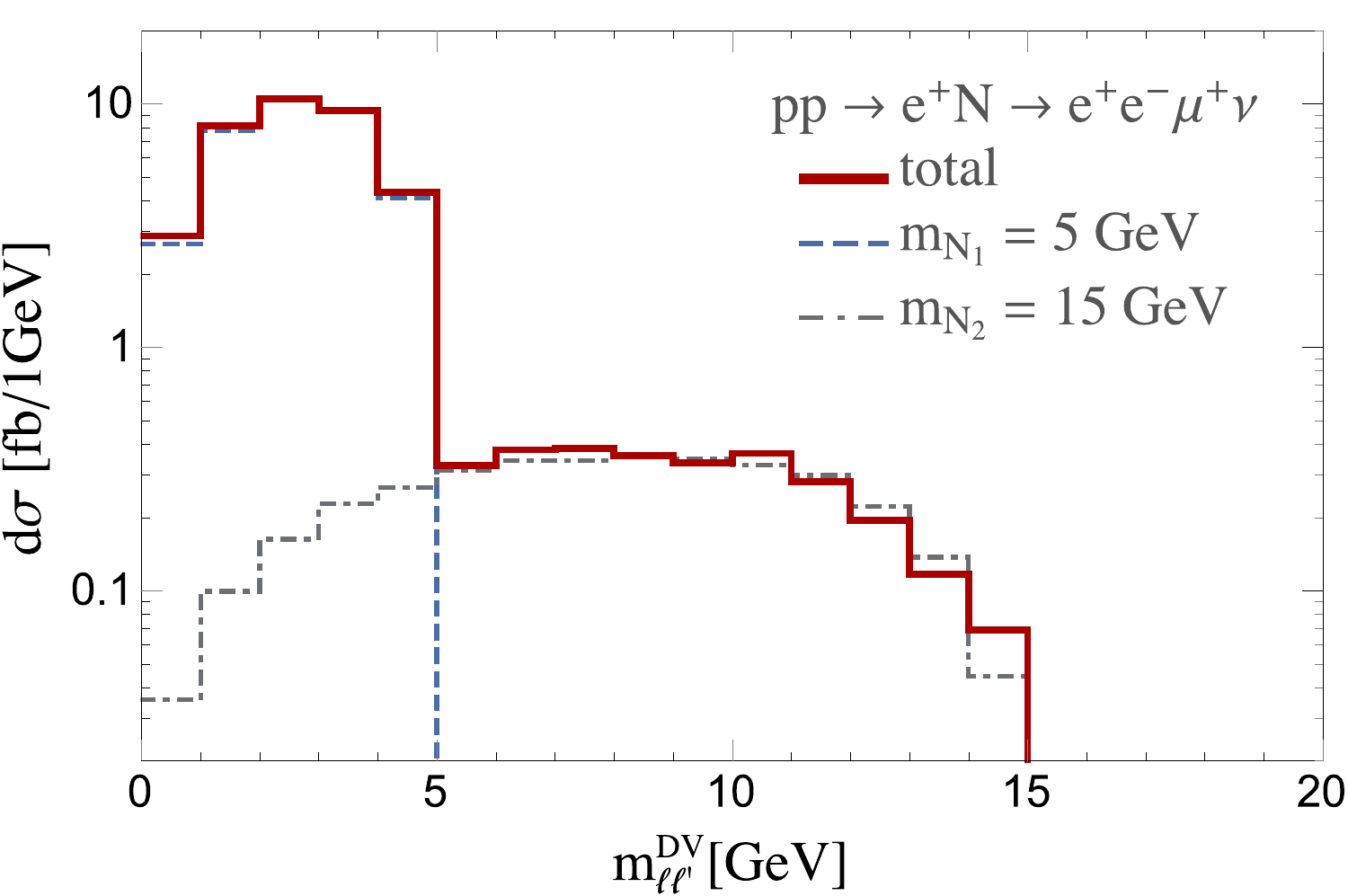}
\includegraphics[width=.485\textwidth]{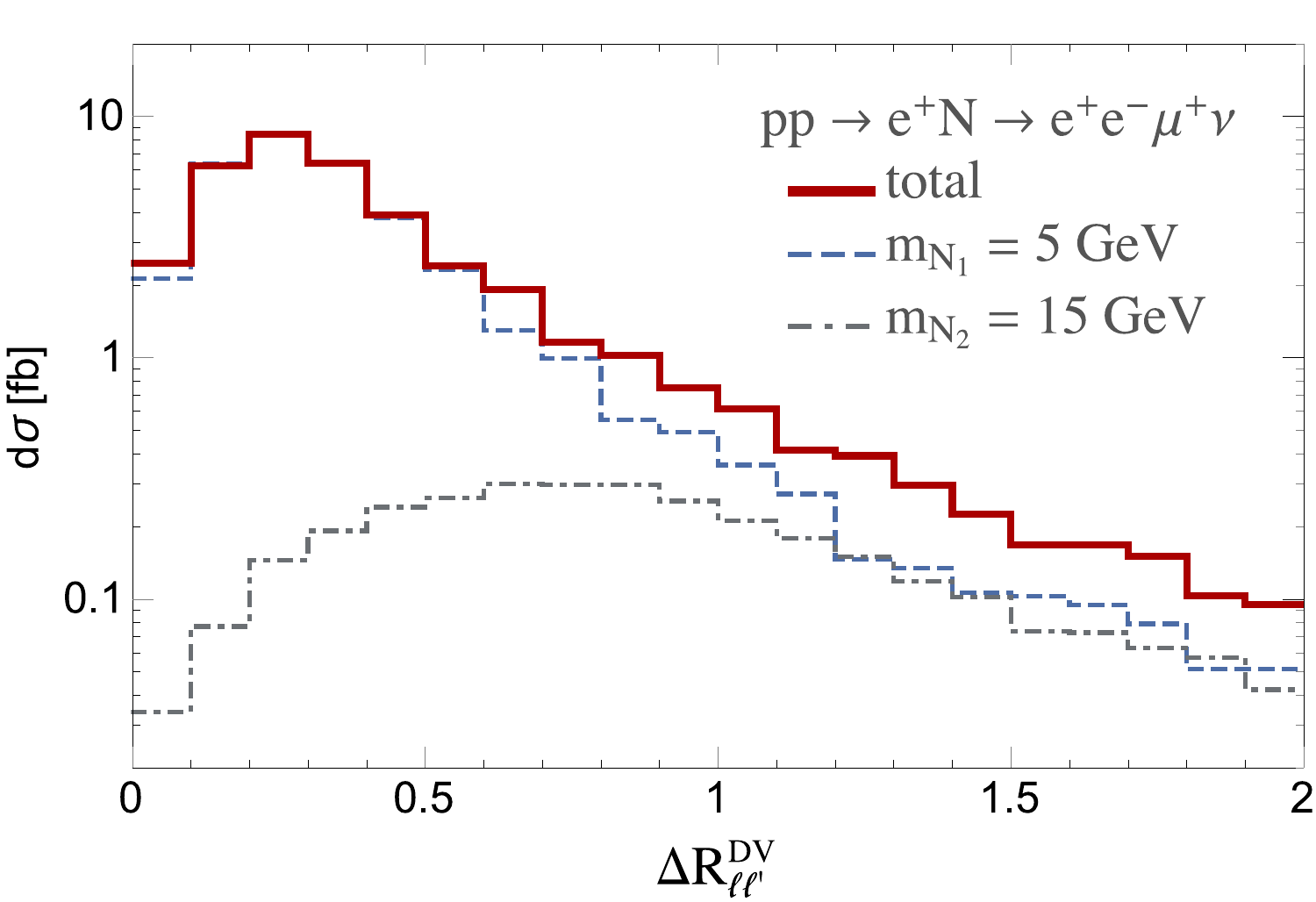}
	\caption{Invariant mass $m_{\ell\ell'}^\text{DV}$ (left) and $\Delta R_{\ell\ell'}^\text{DV}$ (right) distributions of the two charged leptons in the DV for a 3+2 model with $m_{N_1}=5$~GeV, $m_{N_2}=15$~GeV, $|V_{eN_1}|^2=10^{-5}$ and  $|V_{eN_2}|^2=10^{-6}$. Dashed lines are for the 3+1 model for each HNL, while red is the total in the 3+2. Cuts on the prompt lepton of $p_T^{e^+}>25$~GeV and DV condition of 1 mm $<\ell_{\rm DV}<$ 1 m and $z_{\rm DV} <$ 300 mm.}\label{distributions_2HNL}
\end{center}
\end{figure}

On the other hand, if a positive signal is found, one would be interested in disentangling the $\mathcal N$ HNL hypothesis from the single one.
We show in Fig.~\ref{distributions_2HNL} two differential distributions that could help to differentiate the $\mathcal N=2$ from the $\mathcal N=1$ case.
This particular example is for the process $pp\to e^+ N_i$ with $N_i\to e^-\mu^+\nu$ and shows the invariant mass $m_{\ell\ell'}^\text{DV}$ (left panel) and $\Delta R_{\ell\ell'}^\text{DV}$ (right panel) of the two charged leptons originating at the DV.
Dashed lines represent the contribution of each HNL to the total red distribution that would be observed.
It is particularly interesting in the case of $m_{\ell\ell'}^{\rm DV}$, since the individual distributions show edges at $m_{N_i}$, which translates to a {\it kink-like} distortion in the total spectrum. The distribution of $\Delta R_{\ell\ell'}^\text{DV}$ is as expected in this case, with a small deviation from the single HNL hypothesis at large  values for $\Delta R_{\ell\ell'}^\text{DV}$.

\section{Displaced Vertices in Other Standard Model  Processes}\label{sec:boosted}

The success of the LHC and detectors like ATLAS, CMS and LHCb have allowed new and more precise  measurements of SM processes.
The embedding of a HNL into the SM  allows for all typical SM processes that  produce $Z$, $W^\pm$ and $H$ bosons to have additional decay  channels.
In the previous sections we discussed the possibilities when a single $Z$, $W^\pm$ or $H$ boson is produced. We now explore additional processes such as:

\begin{itemize}
\item Dibosons,
\item $W^\pm/Z$ + jets,
\item $t$, $t\bar{t}$ with one the  $W^\pm$ bosons decaying via the HNL,
\end{itemize}
considering the case in which a single  gauge boson is decaying via a HNL.  
As before, depending on the relevant variables $m_N$ and the mixing angle  the HNL can be either prompt or long-lived in these SM channels. 
Nevertheless, the different kinematics of these processes may modify the parameter space area where DV happen.

In particular, for these processes the gauge boson can have a large transverse momenta $p_T$ and  for the mass ranges we are considering the corresponding HNL will be boosted. The smoking gun signature would require a single DV as in the analysis presented in the previous section. This implies that the whole effective displacement region shown in Figs.~\ref{DVevents-mNVlN} and \ref{DVevents-inclusive} would be shifted to the right. To clarify, this will allow a specific point in the ($m_N$, $|V_{\ell N}|^2$) plane for which the $N$ decay is prompt  when it is produced via a Drell-Yan process to be long-lived when produced via one of the above mentioned SM processes.
Therefore the precise transition between the prompt and non-prompt decay of the HNL is slightly different in these processes with respect to the  $W^\pm$ and $Z$ decays we explored in the previous section. 

\begin{figure}[t!]
\begin{center}
\includegraphics[width=.49\textwidth]{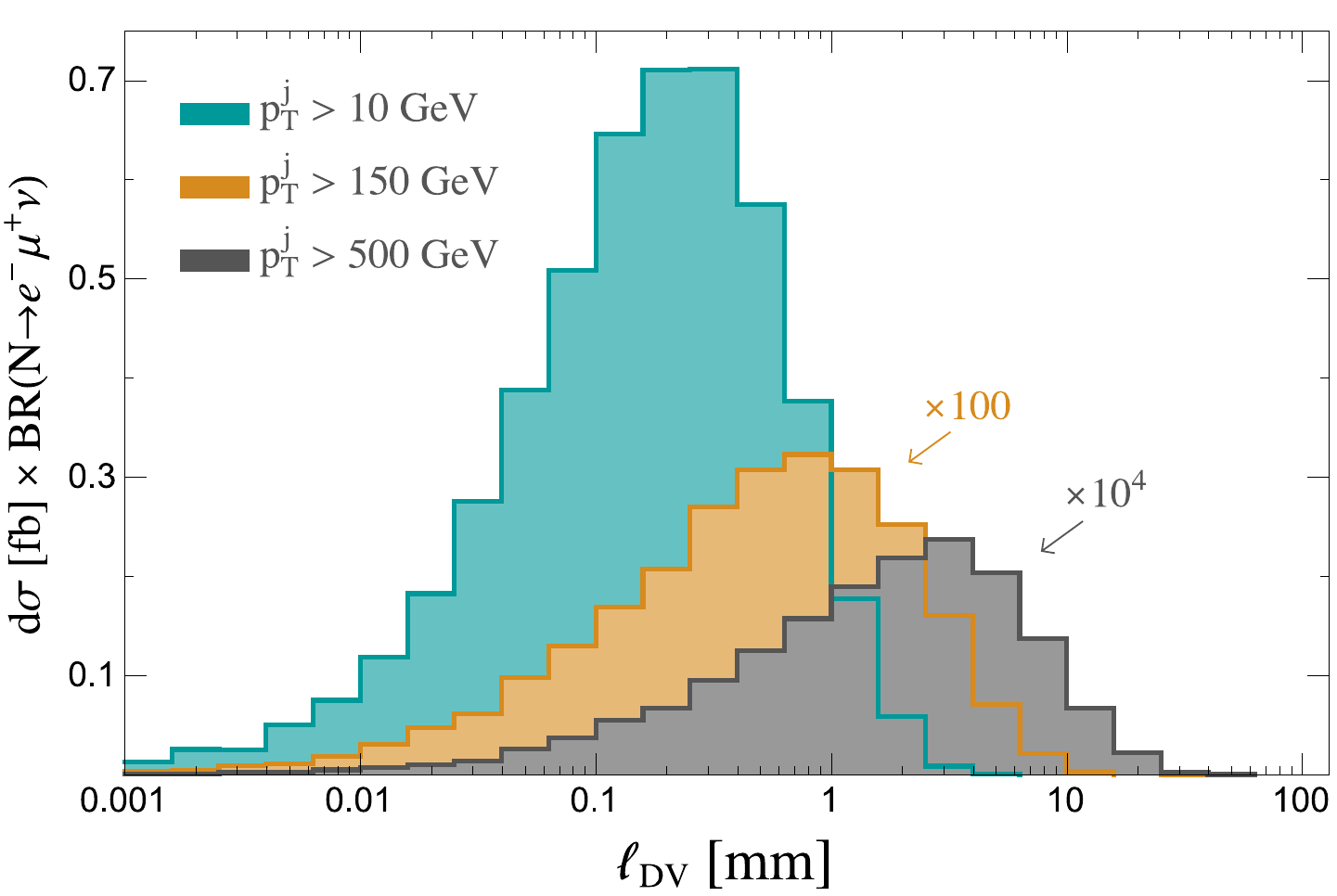}
\includegraphics[width=.49\textwidth]{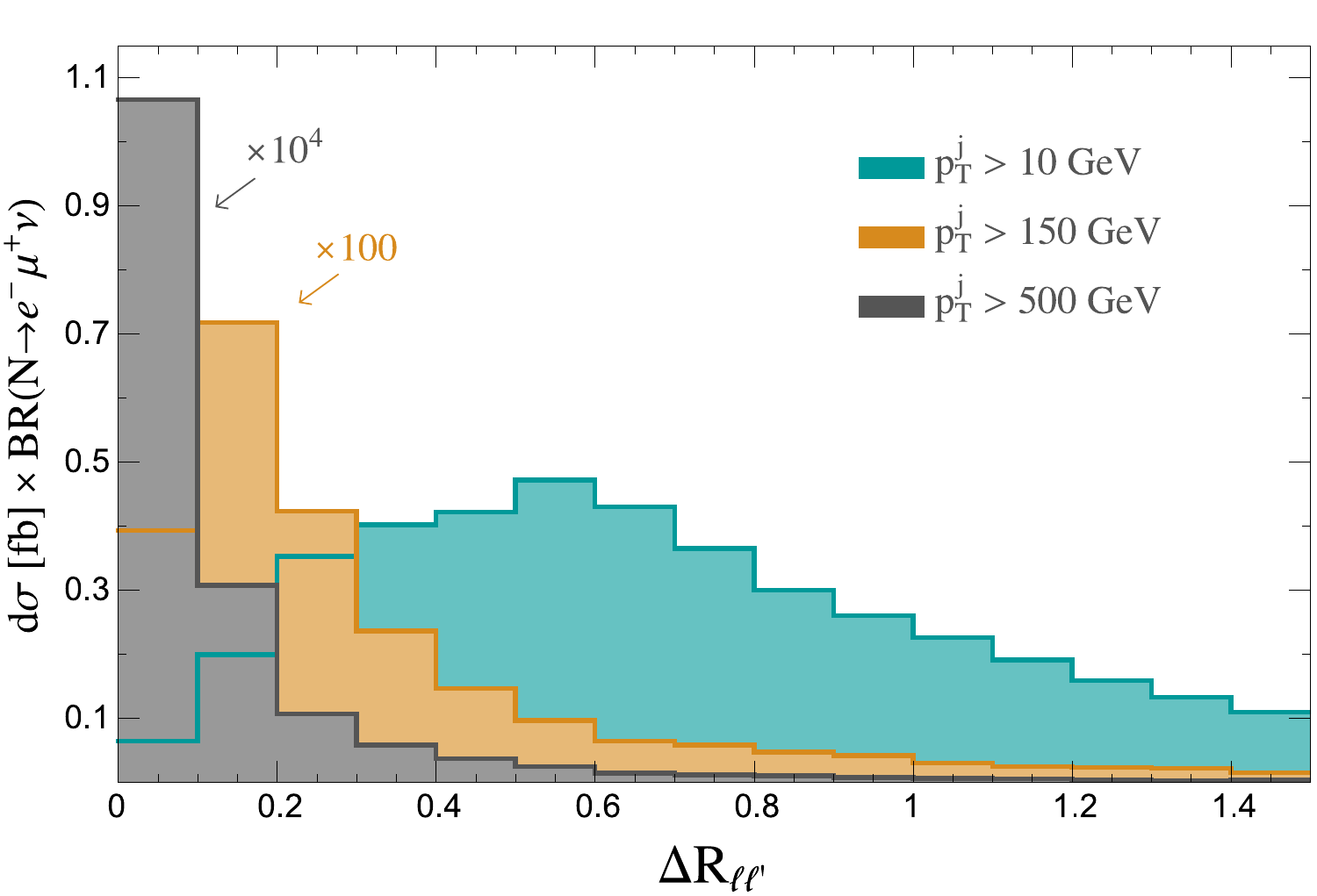}
	\caption{Distributions for transverse displacement (left) and $\Delta R_{\ell\ell'}$ (right) of the two leptons originated at the DV for the process $pp\to e^+ N j$ and $N\to e^- \mu^+\nu$ with $p_T^j>10$, 150, 500~GeV. The neutrino parameters are taken to $m_N=12$~GeV, $|V_{eN}|^2=2\times10^{-5}$  and $V_{\mu N} = V_{\tau N} =0$.}\label{boostedWj}
\end{center}
\end{figure}

We illustrate this effect for the case of the process $pp\to e^+ N j$, where we can induce a boosted topology for the HNL by asking for a high-$p_T$ jet. 
If we select events that have a large transverse momenta of the HNL, then the decay length becomes much larger and significant.  
Fig.~\ref{boostedWj} (left) shows the shift in the transverse displacement from the boosted HNL. 
We see that for a non-negligible shift can be obtained, especially when a very hard jet is present.
Moreover, the decay products of the HNL become more collimated in this case, see Fig.~\ref{boostedWj} (right), which can help to reduce backgrounds when an appropriate cut is implemented.
Nevertheless, these strong cuts severely reduce the number of events and therefore this boosted HNL scenario could be probed only at the  high luminosity LHC.

A more dedicated analysis is needed to study the potential of these SM processes, exploring the different channels mentioned above and considering more energetic hadron and leptonic colliders.
Although experimentally challenging, these new processes would aid in exploring areas of  parameter space that could be difficult to access via other channels.


\section{Conclusions}\label{sec:conclusion}

We have revisited collision phenomenology when heavy neutral leptons (HNL) are added to the particle content of the SM. We do not consider a specific mechanism or model in which these additional neutral fermions are embedded. We have especially focused on the region of parameter space of mass and mixing of the heavy neutral leptons for which the distinctive signature of a displaced vertex (DV) in the inner tracker of a detector is produced when this neutral fermion decays at the LHC.

We have further analyzed the implications on the decay width and the corresponding decay length when different flavor configurations are considered in the production channel via $W^\pm$  and $Z$ gauge bosons. 
While $Z$ and $H$ bosons decays are flavor blind, the $W^\pm$ boson decays are not and the decay length is modified when the HNL couples to more than one flavor with similar (or larger) strength.

Our next step has been to consider an inclusive and flavor blind production in order to determine the precise values of the mass and mixing angles of the HNL in which a sufficient number of events with  measurable DV can occur.  With a dedicated study of the dimuon decay channel case, 
we find that a dedicated experimental analysis could improve the bounds by at least one (two) order of magnitude in the sum of the squared mixings for values of $m_N \lesssim  5 - 15 $~GeV, and  for an integrated luminosity of 300 (3000)~fb$^{-1}$. Nevertheless, the latter sensitivities could be highly improved by lowering the cuts on the lepton transverse momenta. 

We  also show how the distribution for the invariant mass of the leptons (or quarks) produced at the DV can help distinguishing the number of additional heavy neutral leptons that are important in the leptonic mixing matrix.

For the mass ranges of the HNL we are considering for all SM processes in which a $W^\pm$, $Z$ or $H$ boson is produced, their decays are modified by the additional channels that contain the HNL and its subsequent decays. However, the kinematics associated to the HNL varies significantly according to the specific SM process in which it is produced. 
We have shown the impact that a boosted HNL has on the decay length and state that  the analysis of the DV signatures is relevant for these SM processes in the high luminosity phase of the LHC. 

In summary,  we have highlighted in this work the importance of multiple production channels in order to improve the constraints on the region of parameter space for which DV signatures occur when the HNL decays. We clarified the role of flavor and kinematics showing how different SM processes and channels are complementary to establish the existence of a HNL with a mass on the order of a few tens of GeV at the LHC.


\acknowledgments
The authors would like to thank Nishita Desai,  Claudia Garc\'ia-Garc\'ia, Deywis Moreno, Richard Ruiz and Jakub Scholtz  for
fruitful discussions.
We acknowledge partial support from the European Union Horizon 2020
research and innovation programme under the Marie Sk{\l}odowska-Curie: RISE
InvisiblesPlus (grant agreement No 690575)  and
the ITN Elusives (grant agreement No 674896).
NB and ML are also supported by the Universidad Antonio Nariño grants 2017239 and 2018204.
NB was also partially supported by the Spanish MINECO under Grant FPA2017-84543-P. ML thanks LPT-Orsay and CEA-Saclay for hospitality during the completion of this work.
In addition to the software packages cited above, this research made use of IPython~\cite{Perez:2007emg}, Matplotlib~\cite{Hunter:2007ouj}, SciPy~\cite{SciPy} and python-ternary~\cite{marc_2018_1220444}.

\appendix
\section{Parametrization of the $\boldsymbol{3+\mathcal{N}}$ Minimal SM Extensions}\label{app:sec:parametrization}
The extension of the SM with one HNL state reflects into three new mixing angles ($\theta_{14}$, $\theta_{24}$, $\theta_{34}$) (active-sterile mixing angles), two extra Dirac CP violating phases ($\delta_{41}$, $\delta_{43}$) and an extra Majorana phase ($\phi_{41}$) -assuming the Majorana character. The lepton mixing matrix is now given by the product of 6 rotations times the Majorana phases:
\begin{align} \label{eq:3+1rot}
U_\nu^{3+1} &= R_{34}(\theta_{34},\delta_{43}) \cdot R_{24}(\theta_{24}) \cdot R_{14}(\theta_{14},\delta_{41}) 
\cdot R_{23} \cdot R_{13} \cdot R_{12}  \cdot \rm diag(1,e^{i \phi_{21}},e^{i \phi_{31}},e^{i \phi_{41}}) \nonumber \\
&= R_{34}(\theta_{34},\delta_{43}) \cdot R_{24}(\theta_{24}) \cdot R_{14}(\theta_{14},\delta_{41}) 
\cdot U_{\rm PMNS}^{4\times4} \cdot \rm diag(1,e^{i \phi_{21}},e^{i \phi_{31}},e^{i \phi_{41}})\ ,
\end{align} 
where $U_{\rm PMNS}^{4\times4}$ is the $3\times3$ PMNS matrix extended with a trivial fourth line and column, and the rotation matrices  are defined as:
\begin{eqnarray} \label{eq:R}
R_{34}\ &=&\ \left( 
\begin{array}{cccc}
1 & 0 & 0 & 0 \\
0 & 1 & 0 & 0 \\
 0 & 0 & \rm cos \theta_{34} &\rm sin \theta_{34} \cdot e^{-i \delta_{43}}\\ 
 0 & 0 & -\rm sin \theta_{34} \cdot e^{i \delta_{43}}& \rm cos \theta_{34} 
\end{array}%
\right), \nonumber \\
R_{24}\ &=&\  \left( 
\begin{array}{cccc}
1 & 0 & 0 & 0 \\
0 & \rm cos \theta_{24}  & 0 & \rm sin \theta_{24}\\ 
0 & 0 & 1 & 0 \\
0 & - \rm sin \theta_{24}& 0 &  \rm cos \theta_{24} %
\end{array}%
\right), \nonumber \\
R_{14}\ &=& \left( 
\begin{array}{cccc}
\rm cos \theta_{14} & 0 & 0 & \rm sin \theta_{14} \cdot e^{-i \delta_{41}} \\
0 & 1 & 0 & 0 \\
0 & 0 & 1 & 0 \\
- \rm sin \theta_{14} \cdot e^{i \delta_{41}} & 0 & 0 &\rm cos \theta_{14} \\
\end{array}%
\right).
\end{eqnarray}

Similarly, the mixing matrix $U_\nu$ can be extended to the $\mathcal{N}=2$  case as
\begin{equation}
 U_\nu^{3+2}=R_{45}R_{35}R_{25}R_{15}R_{34}R_{24}R_{14}R_{23}R_{13}R_{12}
~\rm diag(1,e^{i \phi_{21}},e^{i \phi_{31}},e^{i \phi_{41}},e^{i \phi_{51}}),
\label{eq:mixing}
\end{equation} 
where $R_{ij}$ is the rotation matrix between $i$ and $j$. 
For instance, the rotation matrix $R_{45}$ is explicitly given by 
\begin{equation}
R_{45}=\left(
\begin{array}{ccccc}
1 & 0 & 0 & 0 & 0\\
0 & 1 & 0 & 0 & 0\\
0 & 0 & 1 & 0 & 0\\
0 & 0 & 0 & \cos\theta_{45} & \sin\theta_{45}e^{-i\delta_{45}}\\
0 & 0 & 0 & -\sin\theta_{45}e^{i\delta_{45}} & \cos\theta_{45}\\
\end{array}
\right),
\end{equation}
and likewise for the other matrices $R_{ij}$ (in terms of $\theta_{ij}$ and  $\delta_{ij}$).

\bibliography{biblio}
\end{document}